\documentclass[acmlarge]{acmart}

\AtBeginDocument{%
  \providecommand\BibTeX{{%
    \normalfont B\kern-0.5em{\scshape i\kern-0.25em b}\kern-0.8em\TeX}}}

\acmJournal{IMWUT}
\acmVolume{0}
\acmNumber{0}
\acmArticle{0}
\acmMonth{0}

\usepackage{caption}
\usepackage{subcaption}
\usepackage{_macros}
\usepackage{multirow}

\newcommand{\name}{Eggly}
\begin{document}

\setcopyright{licensedcagov}
\acmYear{2023} \acmVolume{7} \acmNumber{2} \acmArticle{67} \acmMonth{6} \acmPrice{15.00}\acmDOI{10.1145/3596251}

\title[\name: Designing Mobile Augmented Reality Neurofeedback Training Games for Children with Autism Spectrum Disorder]{Eggly: Designing Mobile Augmented Reality Neurofeedback Training Games for Children with Autism Spectrum Disorder}

\author{\href{https://orcid.org/0009-0008-7730-8552}{Yue Lyu}}
\email{yue.lyu@uwaterloo.ca}
\affiliation{
    \institution{University of Waterloo}
    \city{Waterloo}
    \state{ON}
    \country{Canada}
}

\author{\href{https://orcid.org/0000-0002-7705-2031}{Pengcheng An}, \href{https://orcid.org/0000-0003-4972-4870}{Yage Xiao}}
\email{anpc@sustech.edu.cn}
\affiliation{
    \institution{Southern University of Science and Technology}
    \city{Shenzhen}
    \country{China}
}

\author{\href{https://orcid.org/0009-0001-8578-6449}{Zibo Zhang}}
\email{selenaz.zhang@mail.utoronto.ca}
\affiliation{
    \institution{University of Toronto}
    \city{Toronto}
    \state{ON}
    \country{Canada}
}

\author{\href{https://orcid.org/0009-0001-0373-8241}{Huan Zhang}}
\email{h648zhang@uwaterloo.ca}
\affiliation{
    \institution{University of Waterloo}
    \city{Waterloo}
    \state{ON}
    \country{Canada}
}

\author{\href{https://orcid.org/0000-0002-9642-9666}{Keiko Katsuragawa}}
\email{kkatsuragawa@uwaterloo.ca}
\affiliation{
    \institution{National Research Council}
    \city{Waterloo}
    \state{ON}
    \country{Canada}
}
\affiliation{
    \institution{University of Waterloo}
    \city{Waterloo}
    \state{ON}
    \country{Canada}
}

\author{\href{https://orcid.org/0000-0001-5008-4319}{Jian Zhao}}
\email{jianzhao@uwaterloo.ca}
\affiliation{
    \institution{University of Waterloo}
    \city{Waterloo}
    \state{ON}
    \country{Canada}
}

\renewcommand{\shortauthors}{Lyu et al.}
\begin{abstract}
\cite{dudley2014value}
Autism Spectrum Disorder (ASD) is a neurodevelopmental disorder that affects how children communicate and relate to other people and the world around them.
Emerging studies have shown that neurofeedback training (NFT) games are an effective and playful intervention to enhance social and attentional capabilities for autistic children. 
However, NFT is primarily available in a clinical setting that is hard to scale. 
Also, the intervention demands deliberately-designed gamified feedback with fun and enjoyment, where little knowledge has been acquired in the HCI community. 
Through a ten-month iterative design process with four domain experts, we developed \name{}, a mobile NFT game based on a consumer-grade EEG headband and a tablet. \name{} uses novel augmented reality (AR) techniques to offer engagement and personalization, enhancing their training experience.
We conducted two field studies (a single-session study and a three-week multi-session study) with a total of five autistic children to assess \name{} in practice at a special education center. 
Both quantitative and qualitative results indicate the effectiveness of the approach as well as contribute to the design knowledge of creating mobile AR NFT games.
\end{abstract}



\begin{teaserfigure}
    \centering
    \includegraphics[width=1.\linewidth]{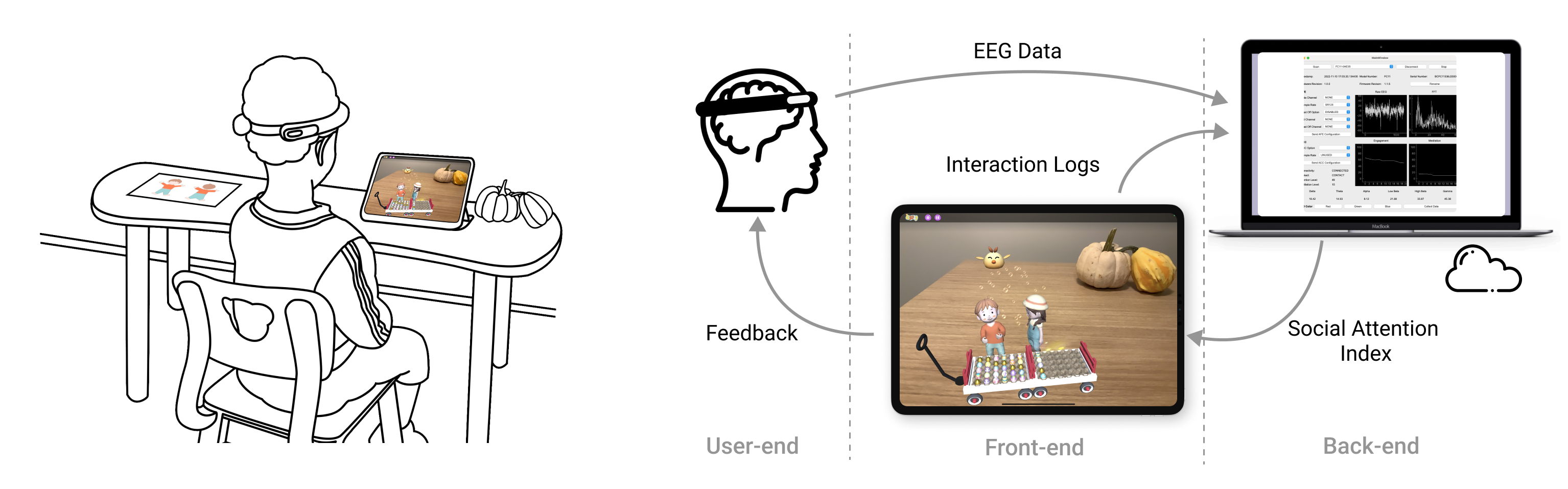}
    \vspace{-10mm}
    \caption{\name{} is a novel mobile augmented reality (AR) neurofeedback training (NFT) game for improving the social and communicative skills of autistic children. Using a portable electroencephalogram (EEG) headband and a tablet, \name{} gamifies feedback in a story where the child needs to help two game characters collect eggs from a bird on a farm by empathizing and focusing on their social collaboration.}
    \label{fig:teaser}
\end{teaserfigure}

\maketitle

\section{Introduction}
Autism spectrum disorder (ASD) is a neurodevelopmental disorder that includes impairments in communication skills and social interactions combined with restricted and repetitive behaviors, interests, or activities \cite{DSM-5}.
It is estimated that 1 in 100 children worldwide suffers from ASD \cite{world-health-organization}.
Autistic children face many difficulties in daily circumstances due to their impaired social communicative functions and restricted interests in people and activities. 
Interventions such as Applied Behavior Analysis (ABA) \cite{Foxx2008-zx} and Early Start Denver Model (ESDM) \cite{Fuller2020-pb}, are important to improve their ability and should be applied early to be successful in areas of life \cite{Rogers1996-pk}.
Recently, \emph{neurofeedback training} (NFT) \cite{Marzbani2016-jx} games have emerged as an effective and playful approach \cite{Holtmann2011-zo, Antle2019}.
NFT games normalize \emph{mu rhythms suppression} of autistic children, by allowing them to observe social interaction scenes that trigger their mirror neuron activity related to imitation and social cognition \cite{Bernier2007-kl, Oberman2007-uy}, and thus improve social behavior for autistic children \cite{Friedrich2015AnEN, Friedrich2014-ns}.
Mu rhythms can be seen as a characteristic oscillation in the EEG signal in 8-13 Hz, which reflect the mirror neuron activity and are abnormally reduced in autistic children \cite{perry2011motor}.
Real-time information about brain activity (\ie, mu rhythms) is suggested as an objective way of measuring children's performance in NFT \cite{Pineda2008-xr, Friedrich2014-ns, Friedrich2015AnEN} and is leveraged to provide gamified feedback (\eg, visually) in a closed loop, reinforcing the children's social brain functioning.

Despite its apparent promise, the high cost and immobility of NFT have been the challenges in scaling up this approach, as most NFT systems are only available through clinics \cite{Birbaumer2013-kp}. 
Recent technologies in consumer-grade electroencephalogram (EEG) headbands could offer an opportunity to deploy NFT games more ubiquitously.
But few studies have looked into developing NFT games based on portable EEG headbands to promote convenience and affordability.
Moreover, NFT needs long-term intervention to take effect; thus, designing gamified feedback to maintain engagement is beneficial \cite{Coben2010-yy}.
While existing research has examined the effects of NFT games (or interventions) \cite{Holtmann2011-zo}, there lacks practice-oriented framework to guide the design of NFT game feedback. 
Last, children may have divergent preferences over game elements (\eg, characters, colors), which require sufficient customization to keep them engaged. 
Augmented reality (AR) could blend the physical and virtual worlds and bring a familiar background or personalized object into the games, thus allowing for a safer and more connected feeling \cite{Mesa-Gresa2018-cr, Chen2016-vp}.
However, little has been explored in integrating AR into NFT game design to enable more affordable, engaging, and personalized experiences. 

To address these challenges, we designed a mobile AR NFT game, named \emph{\name{}}, for preschool children with ASD (30 months to 6 years old), based on a platform consisting of an EEG headband and a tablet.
The game features a story where the child needs to help collect eggs from a bird on a farm by empathizing with two game characters (\autoref{fig:teaser} and \autoref{fig:scenario}).
Through the EEG headband, \name{} captures real-time mu suppression as a measure of the brain function of social skills and interprets it into the \emph{social attention index} \cite{hobson2017interpretation}.
A higher value of the index means a better attention focus and a more ideal mu suppression.
The social attention index is fed back to the child through various means, for example, influencing egg collecting speed, background music tempo, characters' facial expressions, etc.
This forms a feedback loop to strengthen children's brain functioning to self-regulate social attention, which could enhance their social abilities in real life.

We choose the above mobile platform to improve the easiness and affordability of NFT services for autistic children and their families.
To optimize the effect of NFT games, we propose a feedback design framework by investigating the literature and working with four domain experts. The framework depicts different dimensions, \ie, immediate, storytelling, progress, and reinforcing feedback. 
\name{} is designed to fulfill the framework by offering various auditory and visual rewards, which motivate children to improve their performance and engage them in gameplay. 
To accommodate individual differences in social attention level, \name{} dynamically adjusts the game difficulty for each play.
Further, through AR, \name{} allows children to customize in-game characters by coloring on paper sheets, taking pictures of the sheets, and building unique 3D characters from the 2D scans. 
Given the limitations of state-of-art computer vision techniques, Wizard of Oz (WOZ) is employed to implement this feature by manually applying the costumes of the personalized characters.

To evaluate \name{}, we conducted a field study with five ASD children (accompanied by their training caregivers, and sometimes also their guardians) at a special education center. 
The center is configured in a nonclinical setting, and these children were participants already enrolled in specific NFT programs there.
Thus, the environments are familiar to the children, similar to a home setting but with access to staff members' help. 
Further, we conducted a three-week deployment study with one of the children at the same center to gather longer-term outcomes.  
The results of the studies provide both quantitative and qualitative findings about the system's usability, the user experience of the game, and the effects of the novel game mechanisms and features. 
The studies also offer in-depth insights into the design, development, and deployment of mobile AR NFT games as a practical and playful ASD intervention.
In summary, our contributions in this paper include: 
\begin{itemize}
    \item A feedback design framework developed with domain experts to guide the design of NFT games;
    \item A first attempt to integrate AR and NFT in a mobile game, \name{}, that embodies the feedback framework and can be used in everyday (nonclinical) settings with a portable EEG headband and a tablet;
    \item Empirical understanding of ASD children's user experience of \name{} and insights into the design and development of future AR NFT games.
\end{itemize}

\section{Background} \label{sec:background}
In this section, we review research on different interventions targeted at the social skills of autistic children, prior cases of NFT games, and techniques that leverage AR for autistic children's learning and intervention.

\subsection{NFT Games for Autistic Children}
Autism spectrum disorder (ASD) has diverse conditions characterized by difficulty with social interaction and communication \cite{DSM-5}. 
Traditional interventions have been extensively studied and proven to effectively improve their social skills and attention focus \cite{Ferster1964-cz,rogers2016early}, but they usually demand skilled therapists and clinical equipment, while applications have been proposed to monitor these interventions \cite{hayes2004designing}.
New interventions have been explored with new technologies, potentially leading to more effective, accessible, and enjoyable ASD treatment strategies.

Particularly, neurofeedback training (NFT) is an emerging approach for providing on-demand social skills training by allowing
real-time signals of brain activities through feedback (\eg, visual, auditory) to users on a computer interface, thus forming a closed feedback loop to strengthen certain brain activities \cite{Wolpaw2007-wr,Friedrich2014-ns}.
NFT has shown effectiveness in improving social skills for autistic children by targeting mu rhythm modulation \cite{Kouijzer2009-vc}.
By enabling autistic children to watch social interaction sequences, the system triggers their mirror neuron activity (related to abilities of imitation, intention understanding, and empathy \cite{Dapretto2006-wf}) and reinforces the children’s social attentional brain functioning \cite{Oberman2007-uy}. 
However, most commercial-grade NFT systems primarily focus on meditation for neurotypical individuals, where little knowledge exists for designing NFT games for autistic children.

During NFT, children learn to self-regulate their brain activity and maintain it in a particular cognitive state (\eg, attention, relaxation, etc.) \cite{Tan2010}. 
Self-regulation requires continuous effort and a repetitive process; consequently, children easily lose motivation and interest.
To address this issue, researchers incorporate playful gamified designs to maintain engagement, like social mirroring games \cite{Friedrich2015AnEN}. 
For example, in traditional NFT sessions, children are usually required to complete relatively dull, repetitive, and mono-focusing tasks, such as controlling the sizes of bars or paying attention to an auditory signal \cite{Sitaram2017-lr}. 
With the gamified design, tasks are more engaging, for example, making an emoticon smile \cite{Bakhshayesh2011-vz}, a monkey climbing a tree \cite{Bakhshayesh2011-vz}, and a dolphin diving into the ocean \cite{Steiner2014-jo}.

Although gamification adds fun and engagement to NFT, limited practice-oriented design principles are available to guide the design of NFT games.
Also, the high cost and immobility of clinical-based NFT systems are barriers \cite{Luctkar-Flude2019-mo}; more available approaches are demanded by families with autistic children. 
Recently, Antle \etal~\cite{Antle2019} showed the effectiveness of Mind-Full, a mobile NFT application for young children, while not autistic, to self-regulate anxiety.
Based on a co-design approach with four experts, we propose a multi-level feedback framework that could lay the foundation for effective and engaging NFT game design for autistic children.
Our game, \name{}, implements the proposed framework and employs AR to enhance engagement. It also leverages portable EEG and computing devices to address the constraints of traditional NFT.

\subsection{AR Technologies for Autistic Children}
Augmented reality (AR) blends virtuality with reality in real-time, %
offering extra benefits for ASD games. 
Past research has proved that AR can be applied to ASD training, enhancing traditional interventions (\eg, \cite{Charlop-Christy2003-dc,Washington2017-qc}).
One of the benefits of AR is boosting concentration in ASD treatments.
AR can improve the ecological validity of interventions, increasing the sense of presence and engagement for the target users \cite{Waterworth2014-kj,Chen2016-vp}, and improving compliance and positive outcomes \cite{Chicchi_Giglioli2015-mq}.
AR games provide a familiar background while singling out important stimuli in digital format, offering a constrained focus area to reduce the cognitive load of autistic children \cite{Charlop-Christy2003-dc}.
Previous works support this claim and show effectiveness in encouraging children to maintain their focus on social cues \cite{escobedo2014using,AR-Conceptmap,AR-Socialskills}.
Studies in the educational setting also support the positive effect of AR in attracting children's attention \cite{Chang2013-yz,Di_Serio2013-ec}.
This makes AR a desired means to help autistic children learn about social skills.

Another advantage of AR is its portability, as AR products can usually be accessed from mobile devices, including smartphones and tablets. 
For example, AR virtual peers are available on mobile devices \cite{Tartaro2015-sj}, and assistive technologies that help children focus on social interactions can be used with wearable aids (\eg, smart glasses) \cite{Washington2017-qc}. 
While not directly using AR, Marcu \etal~\cite{marcu2012parent} showed success in applying wearable cameras to capture and understand autistic children's context from their perspectives.
Compared to traditional interventions, AR provides an additional convenient and flexible treatment form, especially in home settings.

Finally, AR offers opportunities for customization. 
Children with ASD range widely in their interests and obsessions \cite{Baron-Cohen1999-gj}, and incorporating children's interests into the game could boost their engagement.
Past research has explored scanning children's favorite toys into AR to allow interaction with the toys in a digital space \cite{hodhod2014adaptive}. 
Enabling children to customize game characters can increase their identification with the characters \cite{Kim2015-km}.
Research has found that allowing users to develop game characters (\eg, changing clothes, skin color, etc.) is essential for creating an immersive environment, as users can develop emotional connections \cite{Dickey2006-xt,Mercado2019-oa,Baio2018-il}.

Despite the many benefits of AR-infused ASD treatments, few attempts were made to bridge AR with NFT, an effective yet relatively new intervention type that needs further exploration. 
Our study combines NFT games with AR elements to leverage their advantages for engagement, ecological relevance, and customization. 
Namely, \name{} uses the background of the child's tabletop captured by the camera as the game scene. It also allows children to colorize the costume of the characters on a paper sheet and later input them into the game as 3D characters.

\section{Usage Scenario of \name{}} \label{sec:scenario}
A usage scenario is provided here to afford an overview of the game before diving into the design process and technical details of \name{}.
\name{} narrates a story where a boy and girl collect bird eggs together during a busy harvest season.
The boy collects laid-dropping eggs and hands them to the girl, who places them on a cart trailer.
The child needs to self-regulate their social attention by trying to maintain focus on the characters' interactions in the game scene so that two carts of eggs can be collected as quickly as possible.
Suppose that Jodi, a five-year-old autistic girl, will play \name{} with her mother.

\begin{figure}[tb]
    \centering
    \includegraphics[width=0.8\linewidth]{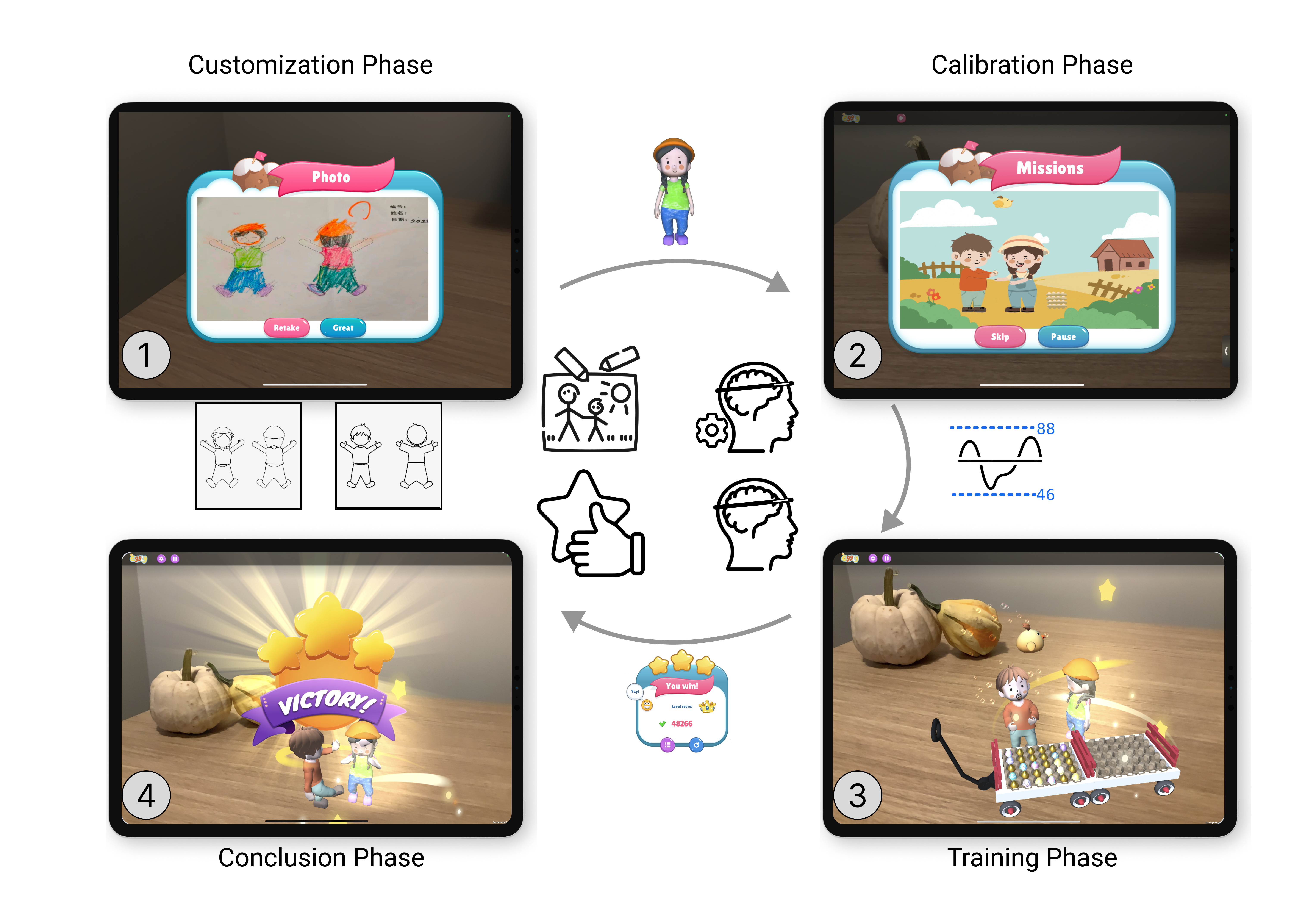} 
    \vspace{-7mm}
    \caption{
    (1) Customization Phase: A picture of a colored sheet is taken to customize the character in the game.. (2) Calibration Phase: An introductory video explains the game story and tasks while determining the customized thresholds. (3) Training Phase: playing \name{} in AR to collect eggs and control the bird's flight height using the thresholds, with performance being logged. (4) Conclusion Phase: Upon completion of the game, a victory scene is displayed, reporting the overall performance.}
    \label{fig:scenario}
\end{figure}

\textbf{Customization Phase.}
Jodi is first provided with paper sheets printed with the outlines of the game characters, the boy and the farmer girl (\autoref{fig:scenario}(1)).
She picks the boy, names him Dudu, and colors Dudu's clothes for her own game session using her gel pen.
Then, Jodi launches \name{} on her tablet. 
From the main menu, she clicks the camera icon to scan the paper sheet she just colored, and submits the picture of Dudu (\autoref{fig:scenario}(1)).
The system transforms Dudu from 2D drawing to a 3D live character to help the farmer collect eggs (\autoref{fig:scenario}(3)). 
Meanwhile, Jodi puts on the headband with assistance from her mother and gets ready to play the game.

\textbf{Calibration Phase.}
Jodi's brain signals are collected in real-time via the headband while she is watching a cartoon introductory video explaining the game story (\autoref{fig:scenario}(2)).
She is then encouraged to attend to the game and the social interactions between the characters to get familiar with the game mechanisms.
She finds out that the more she focuses on the characters' actions, the higher the bird flies, the faster the eggs are being laid, as well as the smoother the characters' collaboration is (see \autoref{sec:feedback-mechanisms} for details).
Based on Jodi's social attention index (a higher value reflecting a better performance),
the system calculates two adaptive thresholds, which optimize the game difficulty for her (see \autoref{sec:adaptive-thresholds} for details). 

\textbf{Training Phase.}
Jodi places her tablet on her little desk and then taps the screen to position the game scene in AR, making the characters and other objects appear to exist on her desk (\autoref{fig:scenario}(4)).
By focusing on the social interaction exhibited in the game, Jodi receives various changes in the game elements as feedback to reward and encourage her. 
The first bird flies sloppily low, and the background music is slow because Jodi lacks attention.
As Jodi focuses more on the scene, the bird flies higher and lays eggs faster, including some colorful ones.
When Jodi keeps her performance, a golden egg is laid and Dudu shows a surprised face.

\textbf{Conclusion Phase.}
After helping collect two carts of eggs, Jodi receives a score out of 100 and stars out of three regarding her overall performance.
This report is also recorded and is useful for her mother to keep track of Jodi's NFT process.
The game then displays a celebration scene where the characters dance together (\autoref{fig:scenario}(e)).
Jodi also intimates the characters' movements and feels happy about completing the task.

\section{Design} 
We employed a co-design approach to develop \name{} by working with domain experts of autistic children.
In the following, we describe this co-design process and then summarize the derived design principles and framework.

\subsection{Co-design Process} \label{sec:co-design}
The co-design process involves four NFT domain experts (E1 - E4) with extensive experience in NFT yet different skills, to bring diverse perspectives and enrich the understanding of the design.
E1 is a research scientist in cognitive psychology at a tech company.
E2 is a professional designer of gamified interfaces for autistic children.
E3 is a university researcher specializing in EEG-based neuroscience and NFT.
E4 is a caregiver from the special education center. 
The entire design process lasted for ten months, during which we closely communicated and collaborated with the experts via various forms, including questionnaires, interviews, co-design sessions, and remote online discussions. 
We did not involve autistic children in the co-design because their limited communication and social skills. 
Instead of parents, we involved E4 who is the actual person accompanying the children in their NFT sessions.
However, we collected the feedback of one child and one parent in the initial system evaluation stage to fine-tune our design.
We now report the five stages of co-design and our gained insights.

\subsubsection{Requirement Gathering (two months).}
In this stage, we aimed to learn about ASD and autistic children, understand current practices and challenges of NFT, and identify principles for developing effective NFT games.
We conducted a literature survey (see \autoref{sec:background} for details) to identify the symptoms \cite{DSM-5} and needs of autistic children \cite{Matson-treatments-ASD}. 
We also explored existing NFT technologies \cite{Friedrich2014-ns, NFT-Behaviour-Pineda, Pineda2008-xr, Kouijzer-EEG, Kouijzer2009-vc, Friedrich2015AnEN, NFT-Spatial-Attention} and their impact on training social skills, as well as works in HCI that focus on design opportunities and challenges \cite{Washington2017-qc, Escobedo2012MOSOCOAM, Benssassi2018,Silva2014-jz,Hong_2013}. 
We carried out two semi-structured interviews with E1 and E2, which led us to choose an easy-setup mobile platform to broaden access to NFT and design for AR games to promote engagement.

\subsubsection{Early Prototype Iteration (two months).}
Following experts' advice, we brainstormed and created three different games in low-fidelity prototypes (\eg, sketches, storyboards, and mock-ups), which all aimed to involve social collaboration tasks regulated by the brain activity to form a feedback loop. 
They also targeted employing AR that could be enabled by a simple mobile phone or tablet.
Next, we presented the low-fidelity prototypes to E1-3 individually and collected their comments in co-design workshops.
The initial design of \name{} was selected, as \pqt{It has the most obvious and straightforward social interaction, characters helping each other and they're facing each other.}{E3}
Moreover, with the help of our experts, we concertized the design objectives for NFT feedback, which were summarized as a feedback design framework exhibiting multiple levels and modalities (\autoref{sec:feedback-framework}).

\subsubsection{System Development (four months).}
We developed a working prototype of \name{} as a tablet application along with a back-end server to display and process the EEG data from the headband.
Through frequent consultation with E1-3, we designed the game elements to promote social collaboration, including two 3D human characters (with obvious body movements) and several facial expressions (aligning with the social collaboration). 
Further, we realized the feedback design framework and detailed implementation of \name{} will be introduced in \autoref{sec:system}.

\subsubsection{Initial System Evaluation (one month).}
Using the working prototype, we conducted a pilot study with one high-functioning autistic child (a six-year-old boy), accompanied by his mother and E4 at the special education center. 
We observed how the child played \name{} and captured his social attention index. 
We asked his mother and E4 to complete a questionnaire regarding their experience, followed by an in-depth interview. 
Overall, they appreciated the game and mentioned that the child was notably more interested in \name{} compared to 2D NFT games.
Based on the comments, we refined the game elements corresponding to the feedback, such as more exaggerated facial expressions, obvious animations, and enriched sound effects.

\subsubsection{Field Studies (one month).}
Using the refined prototype, we conducted a field study with five autistic children and a three-week deployment field study with one child

Details about the study setups and results will be described in Sections~\ref{sec:field-study} and \ref{sec:longterm-study}.

\subsubsection{Resulting Expert Insights.}  
Through the above design stages, we transcribed all our interview sessions with the experts and conducted a continuous thematic analysis of the results. 
For each round of analysis, two of the authors coded the comments independently, then grouped the comments into themes, next conducted discussions to reach an agreement, and finally reported back to the entire research team for further refinement. 
As the feedback mechanism is central to NFT games, we paid special attention to related comments.
Most of the insights were derived from the requirements gathering and early prototype iteration, with some addition in the other stages.
Below we summarize all the key experts' insights obtained from this co-design process. Detailed comments to support these insights are presented in \autoref{appendix:insights}.

\begin{enumerate}[leftmargin=8mm, label=\textbf{I\arabic*}]
    \item[\textbf{I1}] Provide an easy-to-access, cost-effective, and flexible training environment 
    {to offer a safe and controlled setting for autistic children.}
    \item[\textbf{I2}] Offer positive feedback (reward) on the performance gain rather than the absolute performance \rev{to accommodate the limited and wide-ranged social ability of autistic children.}
    \item[\textbf{I3}] Employ multimodal gaming feedback with the visual channel as the primary for its effectiveness \rev{to grab autistic children's attention}.
    \item[\textbf{I4}] Integrate the social attention index dynamically and smoothly into the game scene \rev{to make it easily-understood and avoid distraction for autistic children.}
    \item[\textbf{I5}] Exhibit social collaboration actions in the game that would happen in real life \rev{for autistic children to imitate and apply in their daily lives.}
    \item[\textbf{I6}] Design human-like characters with apparent social gestures and body movements \rev{to allow autistic children to better connect to the real-world}.
    \item[\textbf{I7}] Provide interactive and less-realistic facial expressions \rev{to combat the difficulty of autistic children in recognizing and imitating them.} 
    \item[\textbf{I8}] Exclude textual dialogue in the scene as it interferes with the trigger of mirror neurons \rev{of autistic children.}
    \item[\textbf{I9}] Indicate the performance levels via
    social interactions in the game \rev{for keeping engaging autistic children}.
    \item[\textbf{I10}] Indicate the gameplay progress in the designed social scene \rev{to ease the expectation from autistic children}.
    \item[\textbf{I11}] (Re)enforce the positive feedback to keep engaging and interesting \rev{ autistic children based on their unique sensory and attentional profiles.}
    \item[\textbf{I12}] Leverage AR and 3D graphics to enhance motivation and engagement as well as build real-world connections. %
    \item[\textbf{I13}] Consider customization for children with different severity in ASD.
    \item[\textbf{I14}] Bring children’s familiar/personal objects into the game \rev{to close the gap between the virtual and real worlds} while providing variety across sessions.
\end{enumerate}

\subsection{Design Principles} \label{sec:design-principles}

Based on the obtained insights and the literature, we consolidated the following design principles for NFT games.
The final prototype of \name{} can be seen as an embodiment of these principles, allowing us to investigate our approach to integrating AR into NFT games in a mobile setting for autistic children.

\subsubsection{D1: Lower the barriers for NFT games with more ubiquitous access and easier setup \rev{for families with autistic children}.}
The use of NFT in treating autism has shown promise, but limited accessibility due to factors such as immobility and lack of resources (\eg, clinics, physicians, and infrastructures) makes it difficult to scale. 
To have a broader impact, NFT games should be designed by emphasizing convenience (\textbf{I1}), allowing autistic children to ubiquitously access the intervention with an easy setup.
Previous studies have shown that AR can be used for ASD interventions \cite{Escobedo2012MOSOCOAM, Washington2017-qc}. 
We have taken a first step by combining AR and NFT in the design of \name{} as a low-barrier mobile platform, which can potentially save time and money for families with autistic children.

\subsubsection{D2: Foster multilevel and multimodal game feedback according to \rev{autistic} children's brain activity.}
Current NFT games often employ basic feedback designs, such as variations in 2D graphics \cite{Mihara2012-mh, Mihara2013-of} or pitch of sound \cite{Hinterberger2004-mb}, which lack diversity and fail to sustain attention from autistic children, leading to decreased engagement and reduced NFT performance \cite{Kleih2010-by}.
Our experts advocated for a more varied and fine-grained feedback design to enhance autistic children's engagement in NFT games, which we implemented in \name{}.
First, the feedback should encompass multiple levels to reward different aspects of the performance, such as real-time performance (\textbf{I4}), performance gain (\textbf{I2}), and performance stages (\textbf{I9}) as well as encourage play, such as progress indication (\textbf{I10}), award reinforcement (\textbf{I11}).
Second, the feedback should integrate multiple modalities such as visual and auditory effects (\textbf{I3}), and offer diversity within each level of feedback.

\subsubsection{D3: Embody social collaboration in-game story design while promoting engagement for children.}
As NFT targets addressing the social skills deficits in ASD \cite{Friedrich2015AnEN, Friedrich2014-ns}, the design of the game scene should center around social collaboration \cite{Yang2021-cw}, allowing the child to observe, imagine, and imitate outside of the game (\textbf{I5}).
\name{} features a fun social task, collaborative egg collection, involving several daily actions (\eg, hands up, egg catching, passing, placement).
Additionally, the social interaction design should include apparent gestures and movements (\textbf{I6}) and simplified facial expressions (\textbf{I7}), but exclude text-based dialogue (\textbf{I8}).
To address autistic children's attentional limitation \cite{Adams2012-qb,landry2004impaired}, AR with vivid 3D graphics could be integrated (\textbf{I12}) to relieve their cognitive load \cite{Buchner2022-ci} and guide their focus on relevant stimuli within a bounded viewing area \cite{Charlop-Christy2003-dc, Shipley-Benamou2002-df, Sherer2001-kb}, which enhances the engagement by providing an immersive game scene.

\subsubsection{D4: Allow for personalizing game elements, difficulty, and environment to suit individual children.}
Current NFT games do not account for differences in social attentional capacities among autistic children \cite{Baron-Cohen1988-rc, Dawson2004-vl, Kasari2001-qk}.
Thus, NFT games should be customizable to serve a broader population (\textbf{I13}), such as adapting to each child's social attention level \cite{Whyte2015-fs}; so \name{} employs adaptive thresholds in each session.
In addition, connecting game elements with children's personal objects or creations could increase motivation and engagement (\textbf{I14}) \cite{Dickey2006-xt}. 
Using AR and WOZ, \name{} allows for customizing the game characters with their own colorings.

\subsection{Feedback Design Framework} \label{sec:feedback-framework}

Based on prior experience designing 2D NFT games \cite{Yang2021-cw} and the rich insights gathered during the co-design process, we formulate a feedback design framework for NFT games for autistic children, via the thematic analysis described earlier. 
This framework characterizes four levels of feedback elements including \emph{Immediate} (\textbf{I4}), \emph{Storytelling} (\textbf{I9}), \emph{Progress} (\textbf{I10}), and \emph{Reinforcing} (\textbf{I11}) Feedback.
\autoref{tab:feedback-level} provides a summary of the four levels of feedback, each of which should be realized with multiple modalities to augment the feedback holistically.

\begin{table*}[tb]
\caption{Four levels of feedback for NFT games.} 
\centering
\vspace{-3mm}
\small
\label{tab:feedback-level}

\begin{tabular}{lp{4cm}p{9cm}}

 \toprule
 \textbf{Feedback} & \textbf{Indicator} & \textbf{Example embodiment} \\ 
 \midrule
 Immediate  & Real-time social attention index & Attention meter bar, or simple and direct mapping using game objects \\  
 Storytelling  & Discrete social attention performance stage & Story-related elements, such as collaboration speed or animations of the characters, or other changes in the game scene  \\
 Progress  & Position in terms of completion & Progress bar, or clues integrated with the objects or characters in the game scene  \\
 Reinforcing  & Positive social attention performance & Reinforcers that are favored by children but not directly related to the game story\\
 \bottomrule
\end{tabular}
\end{table*}

\subsubsection{Immediate Feedback.}
This feedback is the real-time indication of the child's social attention index gathered by the headband. 
It immediately visualizes the performance at each time point and thus affords the child's development of self-regulation skills over time \cite{Yang2021-cw}.
From our experts, another key value for it is to help adult facilitators keep track of the child's performance and adjust their assistance accordingly (\eg, verbally motivating the child).
Thus, the immediate feedback often uses simple intuitive representations.

\subsubsection{Storytelling Feedback.}
This feedback corresponds to the child's current discrete stage of performance, 
which should be provided by promoting the storytelling of the game elements related to the characters and the story. %
Examples include the speed of social collaboration, animations of characters' movements, and other changes in the social scene; Yang \etal~\cite{Yang2021-cw} advocated streamlining the narratives.
The higher their performance stage is, the more engaging the storytelling feedback should be, hence creating a positive feedback loop that continuously engages children with the characters and the story.
As the child's performance improves, the storytelling feedback should become more engaging.
As E1 and E2 pointed out, storytelling feedback also needs to be intuitive for children to reason and establish causality between their performance stage and the changes in the game scene. 
 
\subsubsection{Progress Feedback.}
This feedback implies the current progress of game tasks: how much the child has gone and how far it is from completion, providing awareness for both the child and the facilitators. 

This feedback is often indicated by a progress bar on the game interfaces \cite{Yang2021-cw}.
However, E2 and E4 stated that the ideal way to communicate the game progress is to naturally integrate its representation into the game objects or characters, rather than using an extra UI element (which could distract children from the story). 

\subsubsection{Reinforcing Feedback.}
This feedback uses extra reinforcers (\ie, stimuli favored by the children but not directly related to the game story) to further reward their positive social attention behaviors. 
E1 and E3 mentioned that having the right types of reinforcers could effectively facilitate the learning of target children in NFT. 
As E1, E3, and E4 shared, although reinforcers could vary across children, there are certain types of elements commonly favored, such as the visual and sound effects of soap bubbles, vocal cheering, verbal encouragement (could be given by the facilitator), or objects like cars or animals.

\section{\name{} System} \label{sec:system}
In this section, we describe details about the implementation of \name{}, guided by the aforementioned design principles (\autoref{sec:design-principles}) and feedback design framework (\autoref{sec:feedback-framework}).

\subsection{System Architecture}
The entire \name{} system consists of a front-end tablet application for presenting the game in AR, a headband for collecting real-time EEG data, and a back-end server for processing and transmitting the data (\autoref{fig:teaser}). 

\subsubsection{Front-end Application.}
We developed the front-end as a mobile AR application for ubiquitous access (\textbf{D1}). It can be deployed on multiple mobile platforms, while a tablet is recommended for ease of interaction.
We leveraged the ARKit XR Plugin and AR Foundation packages that provide native AR integration with Unity's multi-platform XR API to implement the AR features on iPadOS. 
We built a bi-direction communication channel (\ie, TCP) between the front-end and the back-end to relay all the data (\eg, social attention index, user information).

\subsubsection{EEG Headband.}
We used a consumer-grade EEG headband (BrainCo FocusCalm \cite{brainco}) designed for autistic children because it is comfortable, lightweight, and easy to use (\textbf{I1}).
The headband incorporates sensors to provide a wireless connection to the back-end server and evaluate the child's brain states by examining collected EEG signals.
The headband computes and sends the social attention index to the back-end server in real-time.

\subsubsection{Back-end Server.}
The back-end server was developed using Python, which is responsible for receiving and processing EEG data from the headband, and communicating information (\eg, user interactions, social attention index) with the front-end application via the TCP connection.
We built a pipeline to process, transmit, and store the EEG data as well as other game information for monitoring and further analysis.
The back-end server supports displaying the child's brain activity (\eg, five channels of raw EEG signals captured by the headband) on an interface, enabling facilitators to manage the data and monitor the child.

\subsection{Game Elements Design}
As described earlier, \name{} features social interactions to train autistic children based on a story of a boy and a girl working together to collect eggs from a bird on a farm (\textbf{D3}).
We designed various game elements, including characters, animations, and facial expressions, to create the social scene systematically.
We also prepared rich visual and auditory elements for realizing the feedback design framework, which affords combinations of effects that react to the child's performance to improve their experience and engagement (\textbf{D3}).

\subsubsection{Characters and Models.}
\begin{figure}[tb]
    \begin{subfigure}[b]{0.49\textwidth}
         \centering
         \includegraphics[width=\linewidth]{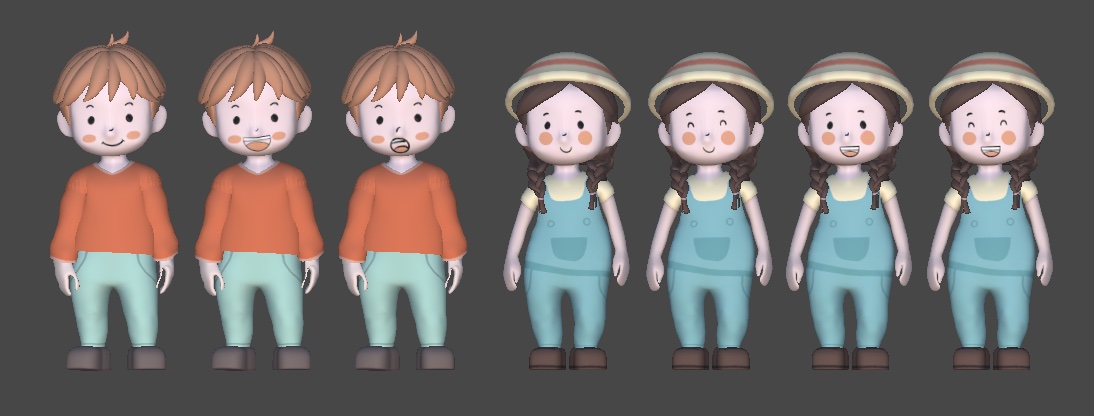}
        \label{fig:facial_expression}
     \end{subfigure}
     \hfill
     \begin{subfigure}[b]{0.49\textwidth}
         \centering
         \includegraphics[width=\linewidth]{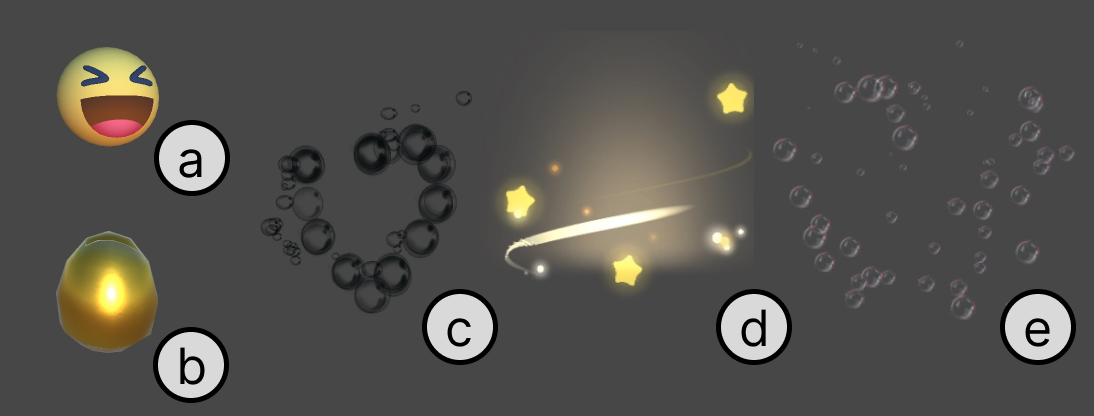}
        \label{fig:visual_effects}
     \end{subfigure}
    \vspace{-7mm}
    \caption{Left) Different facial expressions on characters (left to right): neutral, happy, and expecting faces for the boy; neutral, smiling, happy, and extremely happy faces for the girl. 
    Right) Examples of designed visual effects in \name{}: (a) 3D emoji, (b) golden egg, (c) heart-shaped bubbles, (d) shining stars, and (e) bubbles. }
    \label{fig:expression-effect}
\end{figure}

We chose human-like characters for social interactions in the game to better mimic real-world scenes (\textbf{I6}).
We designed and modelled two 3D characters (\ie, a helper boy and a farmer girl) as shown in \autoref{fig:expression-effect}-Left.
Different parts of the human characters (\ie, hair, shirts, pants, shoes, face) were created with simple strokes and separate texture (UV) mappings, enabling children to easily color every part to customize the characters. 
We created the characters with large hands to emphasize the hand movements, allowing children to focus easily.
Besides the two human characters, we gathered some off-the-shelf 3D models, such as the bird, eggs, egg holders, and cart trailers, to complete our social scene design in \name{}.

\begin{figure*}[tb]
    \begin{subfigure}[b]{0.49\textwidth}
         \centering
         \includegraphics[width=\linewidth]{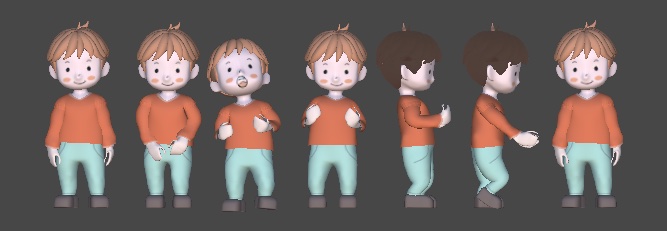}
     \end{subfigure}
     \hfill
     \begin{subfigure}[b]{0.49\textwidth}
         \centering
         \includegraphics[width=\linewidth]{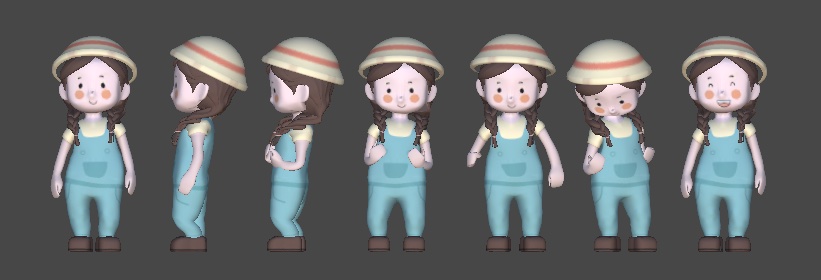}
     \end{subfigure}
    \vspace{-3mm}
    \caption{Different animations/movements of characters. The boy (left to right): base position, hands up, head up, catching eggs, turning with eggs, handing over eggs, and turning back. The girl (left to right): base position, turning to the boy, receiving eggs, turning with eggs, putting down eggs, storing eggs, and turning back.}
    \label{fig:animation}
\end{figure*}

\subsubsection{Animations.}
Characters' animations are critical in our game design to attract autistic children and allow them to mimic social interactions (\textbf{I5}).
We carefully selected animations from the online database Mixamo \cite{mixamo}, and aimed to achieve diverse body movements and gestures to exhibit the vivid social collaboration between the characters (\autoref{fig:animation}).
For the boy, we built movements of catching and delivering eggs (\ie, catching, carrying \& turning, and handing over).
For the girl, getting and storing eggs were implemented (\ie, receiving, carrying \& turning, and putting down).
We also tweaked several animations to make them more realistic for the characters.
The two sets of animations were curated to synchronously work together to demonstrate social collaboration.

\subsubsection{Facial Expressions.}
We designed a variety of facial expressions for our characters, which are essential for encouraging autistic children to match correct facial expressions with certain social scenarios (\textbf{I7}). 
As shown in \autoref{fig:expression-effect}-Left, we created neutral, expecting, and happy faces for the boy and neutral, smiling, happy, and extremely happy faces for the girl, which were used to implement some of the feedback in \name{} (see \autoref{sec:feedback-mechanisms}).
The neutral face was used as the default and the positive expressions were provided accordingly to reflect their social interaction.
For example, the expecting face is shown when the boy heads up and waits for an egg.
A neutral face is used for the boy after catching an egg when the child is in the low-performance stage; however, a smiling or happy face is applied in the medium or high-performance stages, respectively, to reward the child. 
Table~\ref{tab:facial-expression} shows the detailed mapping between the characters' animations and their facial expressions.
\begin{table}[tb]
    \centering
    \small
    \caption{Characters' movements (animations) and corresponding facial expressions. The low, medium, and high in parentheses indicate the performance stages.
    }
    \label{tab:facial-expression}
    \vspace{-3mm}
    \resizebox{\textwidth}{!}{%
    \begin{tabular}{lll}
        \toprule
        \textbf{Character} & \textbf{Animation} & \textbf{Facial Expression(s)} \\
        \midrule
         Boy & Head up for catching & Expecting\\
          & Catching eggs & Neutral (low) or Happy (medium \& high) \\
          & Turning with eggs & Same as above\\
          & Handing over eggs & Same as above\\
          & Turning back & Neutral\\
        \midrule
         Girl & Receiving eggs & Neutral (low), Smiling (medium), Happy (high), or Extremely Happy (>3 seconds in high)\\
          & Turning with eggs & Same as above\\
          & Putting down eggs & Neutral\\
          & Turing back & Neutral\\
        \bottomrule
    \end{tabular}
    }
\end{table}

\subsubsection{Visual and Auditory Materials.}
In addition to animations and facial expressions, we leveraged the Unity asset store to carefully select a set of 3D visual effects, sound effects, and background music.
These materials were designed to work seamlessly with the movements of the characters and objects as well as the designed social interactions. 
The visual and auditory materials are part of the feedback (see \autoref{sec:feedback-mechanisms}) to encourage and guide the child during NFT (\textbf{I3}).
For example, a smiley emoji (\autoref{fig:expression-effect}-Right(a)) shows up when the two characters face each other; also, a coin-clicking sound effect plays with the smiley emoji to reinforce the feedback.

\subsection{Game Customization}
To accommodate individual preferences and differences, \name{} allows for customizing both in-game characters and NFT performance stage thresholds for a game session (\textbf{D4}).

\subsubsection{Coloring Characters.}
The system enables children to scan their personalized 2D coloring sheets of the characters and transform them into 3D in-game characters (\textbf{I14}), as demonstrated in \autoref{sec:scenario}.
If needed, this customization can be performed prior to each game session.
\autoref{fig:scenario}(a) shows the coloring sheet for each character including both sides of the body, and the child could color any number of components. 
While recent advances in computer vision have shown the possibility of converting 2D images to 3D models using deep learning \cite{Zubic2021-nn}, it is still challenging to build a reliable transformation without any artifacts. 
We experimented with several off-the-shelf machine learning models (\ie, Monster Mash \cite{dvorovzvnak2020monster}, PIFuHD \cite{saito2020pifuhd}, Animated Drawings \cite{animateddrawings}), but none of them could stably generate desired results.
While Monster Mash can produce a nice 3D model looking from the front and back, the sides of the characters need much manual fixing.
PIFuHD can efficiently generate models from human photos, but realistic human characters are not ideal for autistic children based on our experts' comments and prior studies \cite{silva2015motivational}. 
Animated Drawings converts hand drawings into animated characters, but only in 2D.
Autistic children's drawings in the wild contain much noise and variety, making the task even more challenging. 
To enhance the user experience, we adopted a WOZ approach by manually integrating digital scans of coloring sheets as textures for rendering 3D characters.
The coloring sheets were created based on 3D model textures, so the digital scans could be easily used as costumes. 
We trimmed the scans to match the texture frame size of the models and improved them (\ie, brightened the costumes, filled in blank spaces) while preserving the child's original coloring, and imported the scans to the game afterward.
This process could take 10-15 minutes for one coloring sheet.
This way, it can produce nearly-perfect 3D characters in \name{}. 
However, future endeavors are needed to make this customization phase faster with automation.

\subsubsection{Adaptive Thresholds.}\label{sec:adaptive-thresholds}
As noted in \autoref{sec:scenario}, \name{} calculates two adaptive thresholds during the calibration phase to adjust the difficulty (\textbf{I13}).
By evaluating the baseline condition without stimuli (feedback), we can determine the thresholds and measure the change in mu suppression with the addition of stimuli in the training phase.
Real-time social attention index is captured through the EEG headband while the child is watching the one-minute introduction to the game.
The system then computes an averaging baseline value $b$ for the child, and two thresholds $[t1, t2]$ are derived based on the baseline: $t1 = \max(LB, b \cdot \alpha), t2 = \min(UB, b \cdot \beta)$,
where $\alpha=0.8$, $\beta=1.3$, $LB = 10$, and $UB = 85$ are empirically determined with the consultation of our experts. 
The two thresholds divide the whole social attention index ranging from 0 to 100 into three performance stages, each corresponding to a set of visual and auditory feedback (\autoref{sec:feedback-mechanisms}) to engage and encourage the child. 
The value 0 represents that children completely lose focus on their observation (no mu suppression of their brain activities)
and 100 means children are fully focused on the observation (a significant mu suppression).
When necessary, the facilitator can designate the thresholds before each session based on their knowledge.

\subsection{Feedback Mechanisms} \label{sec:feedback-mechanisms}
The core of NFT games is the deliberately designed feedback mechanisms that generate a gamified rewarding system to reinforce the children's social brain functioning. 
Guided by the feedback design framework (\autoref{sec:feedback-framework}), we developed diverse feedback elements, together offering multi-level and multimodal feedback holistically in \name (\textbf{D2}).
Table~\ref{tab:feedback} summarizes the feedback implemented in \name{}.

\begin{table*}[tb]
\centering
\scriptsize
\caption{The implemented feedback framework in \name{}. } \label{tab:feedback}
\vspace{-3mm}
\resizebox{0.9\textwidth}{!}{
\begin{tabular}{ll|l}
\toprule
 \multicolumn{2}{l|}{\textbf{Feedback}} & \textbf{Implementation}\\ \midrule
 Immediate & Visual &  Bird flying height directly mapped to social attention index\\
\midrule
 \multirow{6}{*}{Storytelling} & \multirow{4}{*}{Visual} & Slow, normal, and fast body movements for low, medium, and high performances  \\
            &       & Slow, normal, and fast egg laying for low, medium, and high performances   \\
            &       & Facial expressions as listed in Table~\ref{tab:facial-expression} \\
            &       & Heart-shaped bubbles (keeping the high performance stage over 3 seconds)\\ \cmidrule{2-3}
 
            &Auditory&Low, medium, and high background music tempo for low, medium, and high performances\\
\midrule
 \multirow{7}{*}{Progress}  & \multirow{4}{*}{Visual} &  The number of stored eggs on the carts \\
           &        &   Colorful halo above the eggs after collecting a row (5) of eggs \\
           &        &   Shinning stars and light above the tray after collecting a tray (30) of eggs \\
           &        &   Stars after game completion \\ \cmidrule{2-3}

           & \multirow{3}{*}{Auditory}&  ``Woohoo'' sound after collecting a row (5) of eggs \\
           &        &  ``Oh Yea'' sound after collecting a tray (30) of eggs \\
           &        &  Victory drum beats sound after game completion\\
\midrule
 \multirow{4}{*}{Secondary} & \multirow{3}{*}{Visual} &   Randomly-colored eggs \\
  &        &   Golden eggs (none-decreasing performance over 3 seconds) \\
           &        &   Blasting bubbles during eggs being handed over\\
           &        &   Emoji when the characters are facing each other\\
           \cmidrule{2-3}
           & \multirow{2}{*}{Auditory} & Bubble blasting sound\\
  &    &  Coin-clicking sound when the characters are facing each other \\
\bottomrule
\end{tabular}
}
\end{table*}

\subsubsection{Immediate Feedback.}
This immediate feedback, which reflects the (quasi-)real-time social attention index of the child, was directly mapped to the height of the bird in \name{}.   
This was designed to be easily noticed by the child and the facilitator as a direct performance reference (\textbf{I4}).
We designed this indicator that is naturally integrated into the central scene while still remaining effortless to interpret for users.

\subsubsection{Storytelling Feedback.}
We mainly embedded the storytelling feedback in the movements/gestures of the social interactions, because such feedback is given to reflect and reward the performance in discrete stages (\textbf{I9}).
For instance, when the child falls into the low-performance stage, the characters' movements are sloppy and slow.
As they focus more and enter higher performance stages, the bird lays eggs faster, resulting in more active collaboration between characters.
The facial expressions of the two characters also change accordingly.
For example, when the girl gets the egg from the boy, her facial expression is neutral, smiling, and happy, from the low to high-performance stages, respectively. 
More details are shown in \autoref{tab:facial-expression} and \autoref{fig:expression-effect}.
In summary, all the storytelling feedback was designed visually and auditory apparent, allowing children to intuitively establish the causality between their social attention performance stage and the provided feedback.

\subsubsection{Progress Feedback.}
This feedback subtly indicates and rewards player progress (\textbf{I10}) in \name{}.
First, the number of eggs stored in the cart trailer indicates the general progress, where in total, 60 eggs are needed to reach the goal.
During the game, children can effortlessly know how far they are by noticing empty spots on the cart trailer (\autoref{fig:scenario}(d)).
Whenever five eggs (\ie, a row in a tray) are collected, some visual and auditory effects are given to reward such small steps of progress (Table~\ref{tab:feedback}).
A visual effect with stars (\autoref{fig:expression-effect}-Right(d)) rewards the child when the game task is half done (\ie, collecting one cart of eggs), with a voice shouting ``Oh yeah!'' to motivate the child.
Moreover, stars out of three are shown with victory drum beats to indicate the completion of the current game session and the children's overall performance (\autoref{fig:scenario}(e)).

\subsubsection{Reinforcing Feedback.}
\begin{figure}[bt]
    \includegraphics[width=\linewidth]{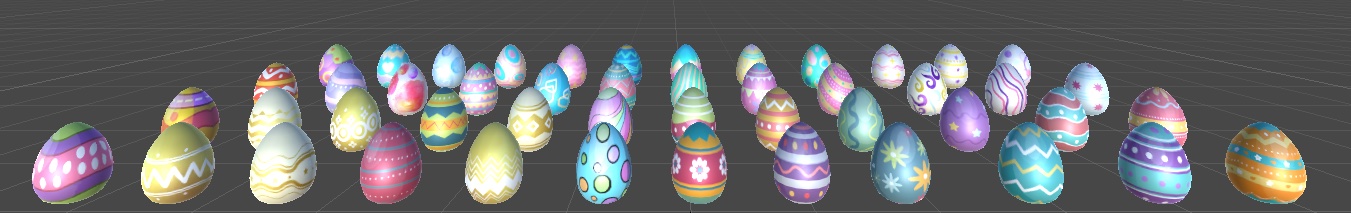}
    \vspace{-7mm}
    \caption{Randomly-colorized eggs laid by the bird to enhance game engagement.}
\label{fig:eggs}
\end{figure}

This feedback in \name{} leveraged familiar or favorite objects of individual autistic children and thus provides an extra level of reinforcement in the reward system (\textbf{I11}).
We implemented the blasting bubbles (\autoref{fig:expression-effect}-Right(e)) as they are commonly liked by autistic children based on the knowledge of our experts.
Also, we implemented the laid eggs in various colorful appearances (\autoref{fig:eggs}), and an emoji face, appearing when the two characters are facing each other (\autoref{fig:expression-effect}-Right(a)).

\section{Field Study} \label{sec:field-study}

We conducted a field study at a special education center to empirically understand how autistic children, their families, and caregivers experience \name{}%
in the real-world context.
\name{} is viewed as a technology probe for investigating the new approach of augmenting NFT with AR and mobile devices.

\subsection{Participants}
We recruited five autistic children with their caregivers/guardians (\autoref{tab:participant-bios}) at the special education center (a non-clinical setting) that provides particular programs (including NFT) to children with ASD and other social skills deficits.
Similar to how their current NFT training was carried out, each child participated in the study accompanied by one caregiver (\ie, staff at the special education center) and sometimes also one of their guardians (\eg, parents) (\autoref{fig:field-study}(a)).
This is because preschool autistic children have limited communication ability and need supervision during the training, as practiced in similar studies \cite{Matson-treatments-ASD}.
Our aim was to evaluate \name{} in a realistic setup similar to typical NFT sessions.
Note that we targeted a specific group of users---preschool children with ASD.
Based on a survey of 79 ASD social-training studies \cite{Matson-treatments-ASD}, the sample size for most experiments is less than five, which is echoed by relevant HCI studies regarding ASD techniques \cite{hayes2004designing, marcu2012parent, Escobedo2012MOSOCOAM}. 
Thus, our study sample size falls within the typical range.
This study was approved by the research ethics office at the related institutions. 

\begin{table*}[tb]
\caption{List of children and caregivers/guardians who participated in the field study.} 
\centering
\small
\label{tab:participant-bios}
\begin{tabular}{lrrrr}
 \toprule
 \textbf{ID} & \textbf{Accompany} & \textbf{Functioning Level} & \textbf{Gender}  & \textbf{Age(year)}  \\ 
 \midrule
P1 & T1, T5 & High & M &  3\\ %
P2 & T2 & Medium & M & 3 \\
P3 & T3 & Low & F& 3 \\ %
P4 & T4 & Medium & M & 6 \\ %
P5 & T5, T6 & High &M &4  \\  %
 \bottomrule
\end{tabular}
\end{table*}

\subsection{Procedure}
\begin{figure}[bt]
     \centering
     \begin{subfigure}[b]{0.33\textwidth}
         \centering
         \includegraphics[width=\textwidth]{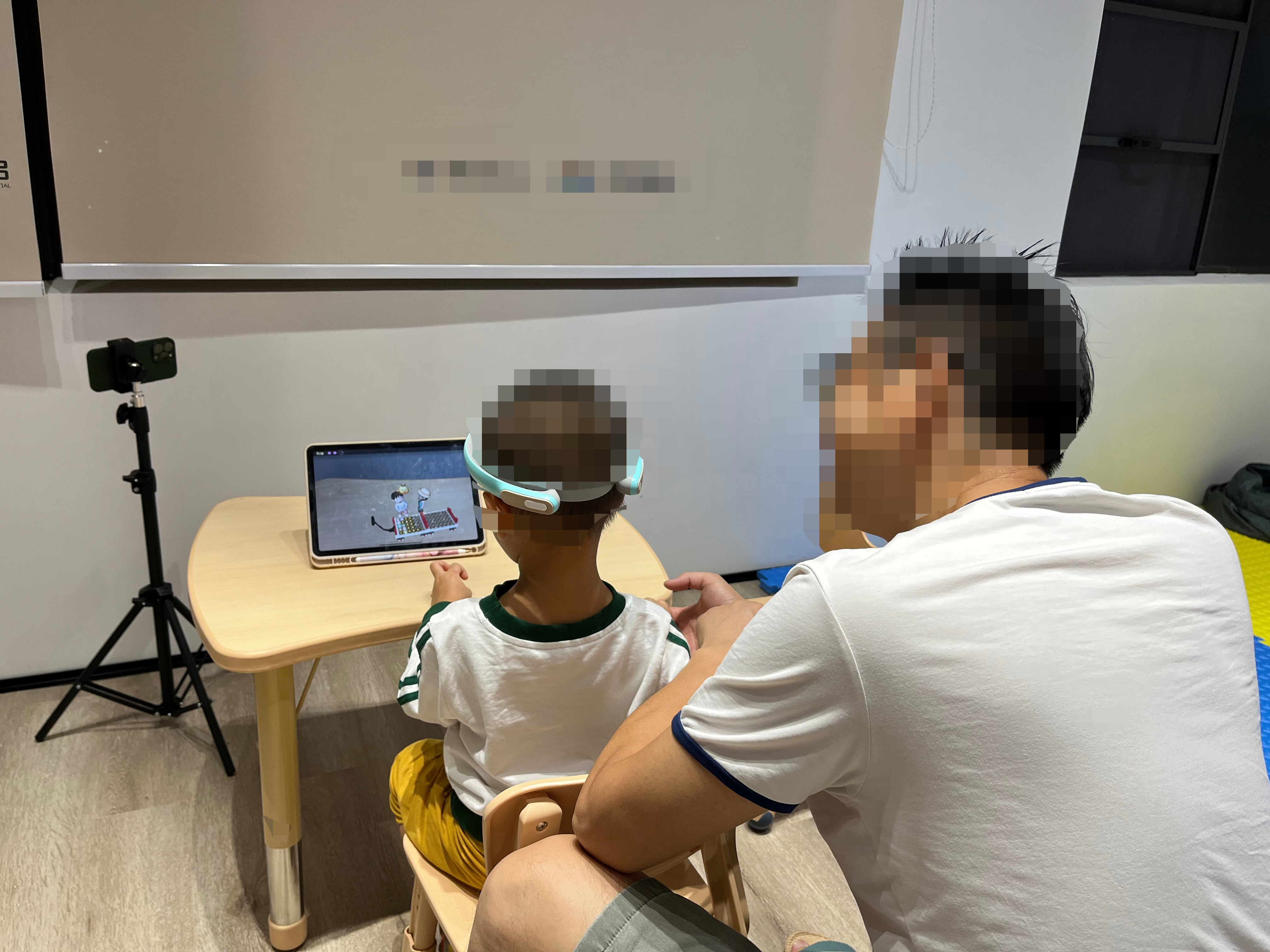}
         \caption{}
         \label{fig:study_site}
     \end{subfigure}
     \hfill
     \begin{subfigure}[b]{0.33\textwidth}
         \centering
         \includegraphics[width=\linewidth]{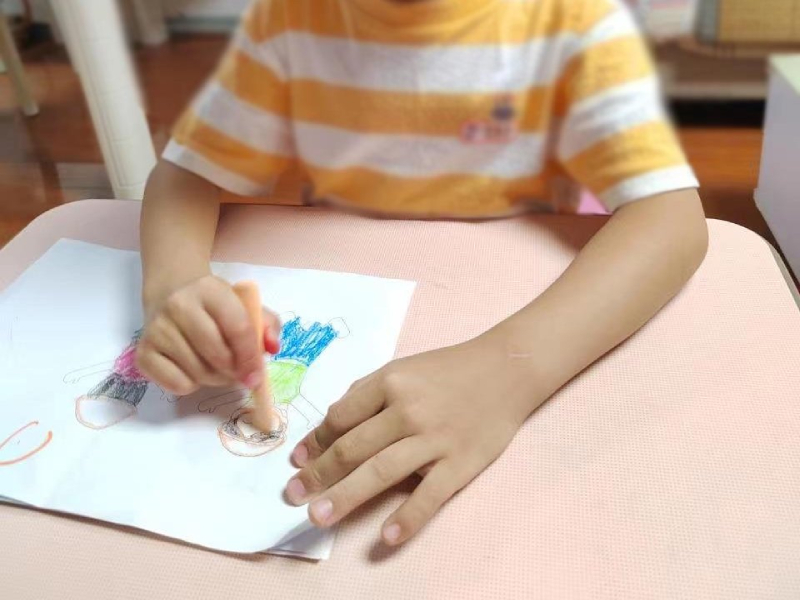}
         \caption{}
         \label{fig:participant_coloring}
     \end{subfigure}
     \hfill
     \begin{subfigure}[b]{0.33\textwidth}
         \centering
         \includegraphics[width=\linewidth]{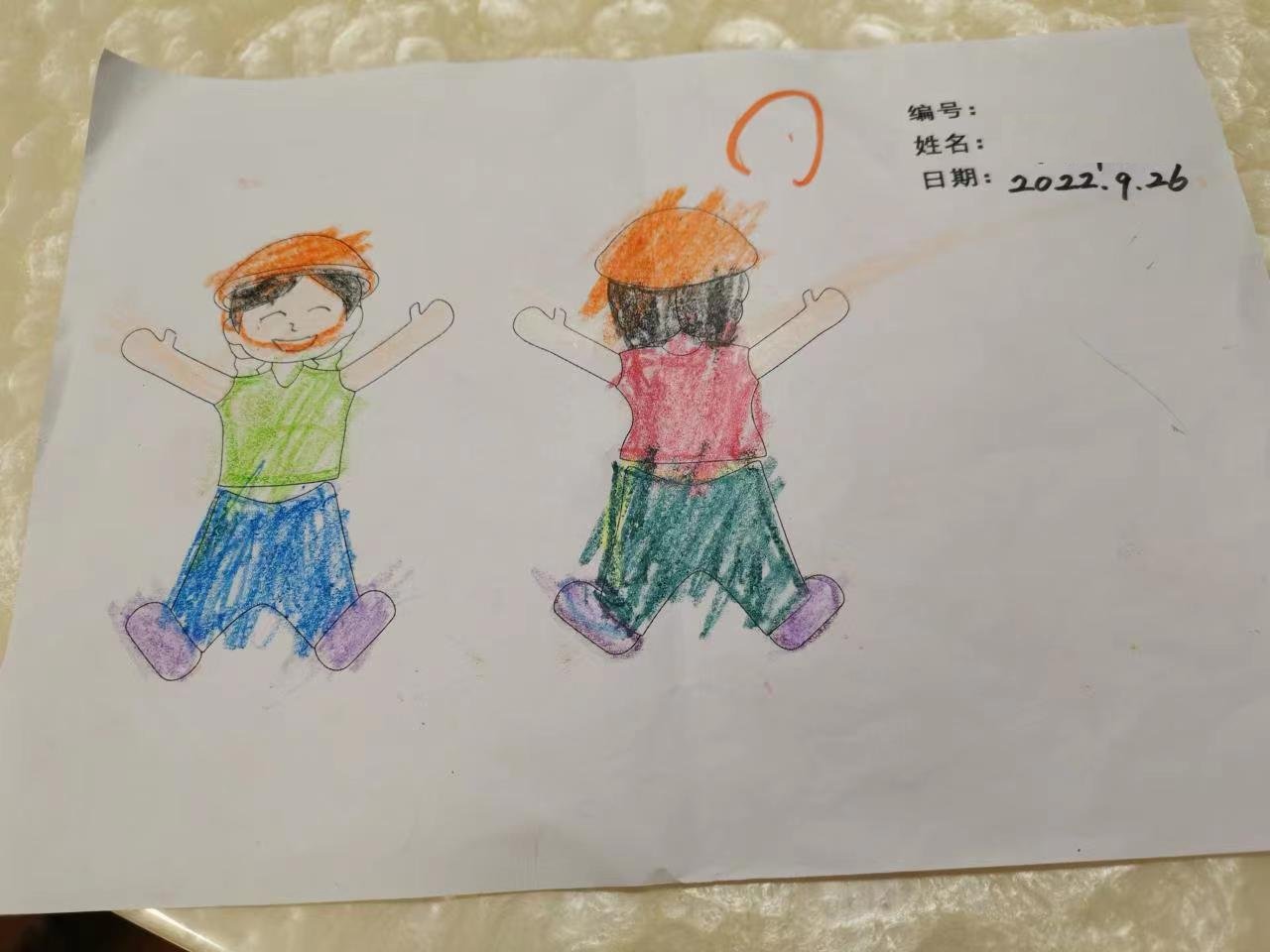}
         \caption{}
         \label{fig:color_sheet}
     \end{subfigure}
     \hfill
     \begin{subfigure}[b]{0.33\textwidth}
         \centering
         \includegraphics[width=\linewidth]{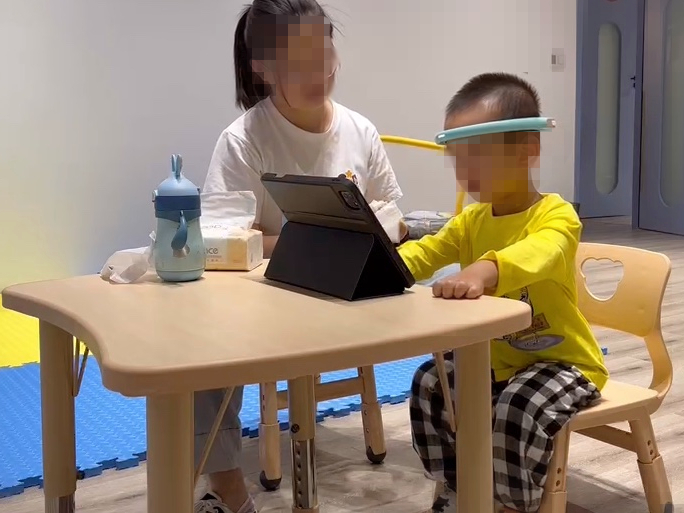}
         \caption{}
         \label{fig:colored_sheet}
     \end{subfigure}
     \hfill
     \begin{subfigure}[b]{0.33\textwidth}
         \centering
         \includegraphics[width=\linewidth]{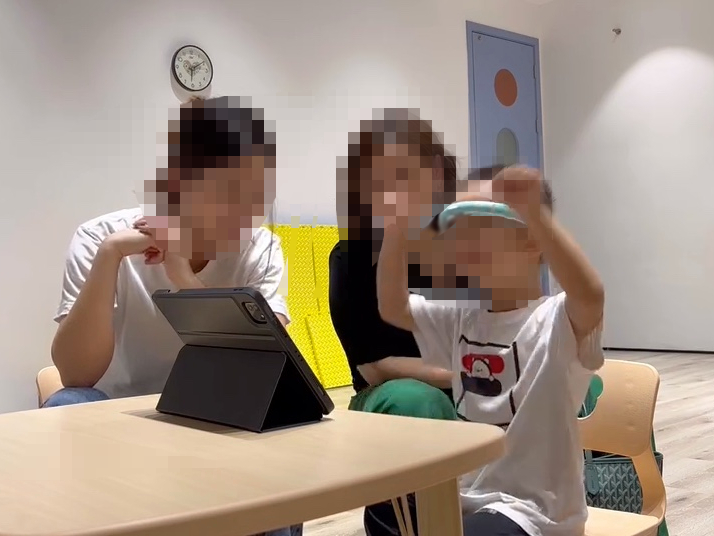}
         \caption{}
         \label{fig:jinlin_cheerup}
    \end{subfigure}
    \hfill
    \begin{subfigure}[b]{0.33\textwidth}
         \centering
         \includegraphics[width=\linewidth]{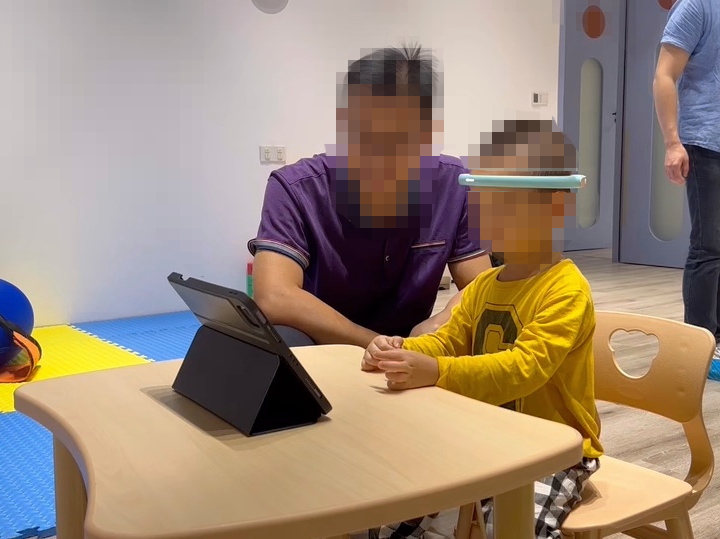}
         \caption{}
         \label{fig:jinlin_with_dad}
    \end{subfigure}\\
    \vspace{-4mm}
    \caption{(a) Study setup at the special education center. (b) A child was coloring the characters during the field study and (c) his finished coloring sheet. (d-f) Gameplay sessions with caregivers/guardians during the studies.}
    \label{fig:field-study}
\end{figure}

After obtaining consent from the guardians and caregivers, the researcher explained the background and process of the study.
Each child was then given the coloring sheets for customizing the two characters based on their preferences (\autoref{fig:field-study}(b,c)).
Next, the caregiver/researcher helped take a picture of the colored sheet, which is then imported to the \name{} system and transformed into 3D characters in the game through a WOZ process.
Meanwhile, the caregiver/researcher supported the child in setting up the tablet and wearing the headband appropriately.
As described in \autoref{sec:scenario}, the child first went through the calibration phase while watching the introductory video.
Similar to current NFT practice, during the gameplay, the caregiver accompanied the child in the whole session and sometimes offer verbal encouragement (\autoref{fig:field-study}(d)).
In the end, the caregivers (and the children's guardians if they attended the session) were invited to fill in a questionnaire, followed by a semi-structured interview to gather their experience and opinions about the game.
Following previous study protocols \cite{marcu2012parent}, we did not interview the children because they have limited abilities in communication, and the caregivers, as the children's regular NFT facilitators, knew the personality and preferences of the children well and were qualified to speak on their behalf.
The entire session was video/audio-recorded for later analysis and lasted about 60 minutes.  
The system also captured gameplay logs with timestamped game events and EEG signals.

\subsection{Result Analysis}
In this section, we report the results of our field study, including the subjective ratings on the questionnaire, EEG logs, and qualitative comments on \name{}.
For the questionnaires and interviews, we collected feedback from six individuals (five caregivers and one guardian). 
In the following, we regard the children as P\# and the caregivers/guardians as T\#, among our participants in the study.

\subsubsection{Questionnaire Responses}
\begin{figure}[tb]
    \includegraphics[width=\linewidth]{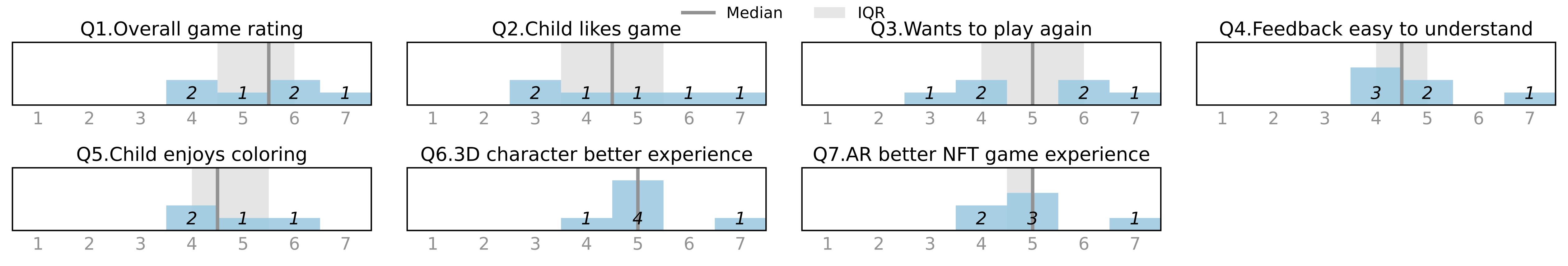}
    \vspace{-9mm}
    \caption{Participants' ratings on the questionnaire, where Q1-4 regard general experience, Q5 regards customization coloring experience, and Q6-7 regard AR experience (1: Highly disagree, 7: Highly agree). }
    \label{fig:questionnaire_responses}
\end{figure}

\begin{figure}[bt]
    \includegraphics[width=0.9\linewidth]{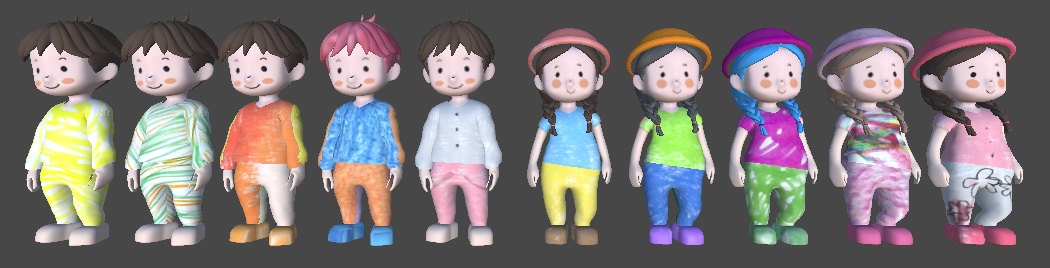}
    \vspace{-4mm}
    \caption{Customized 3D characters generated using WOZ from the paper sheets colored by the children.}
    \label{fig:coloredcharacters}
\end{figure}

The questionnaire was designed to assess different aspects of our approach, such as the overall experience, designed game features, customization process, and AR, as shown in \autoref{fig:questionnaire_responses} on a 7-point Likert scale.
Overall, participants had a positive impression of the game (Q1), wanted to play it again (Q3), and thought the designed feedback was easy to understand (Q4), with the medians all above 4.5 and the majority of the ratings above 4.
As an exception for Q3, P4 seemed to be a little hesitant to play the game again (rated 3), since the child was not familiar with the game theme (farm) and did not understand the game story clearly.
While the median rating on liking the game (Q2) was 4, there was a wider distribution which may be due to children's different needs and preferences for game elements (discussed further in \autoref{sec:qualitative-results}). 
Moreover, the children generally enjoyed the coloring process for customizing the game characters (Q5), demonstrating the benefits brought by this new game feature.
\autoref{fig:coloredcharacters} show some examples of the customized characters by the children.
Participants also suggested that both 3D characters and AR (Q6 and Q7) contributed to a more engaging user experience compared with traditional 2D NFT games, which confirmed our initial design ideas.

\subsubsection{Interaction Logs}
\begin{figure}[tb]
    \includegraphics[width=\linewidth]{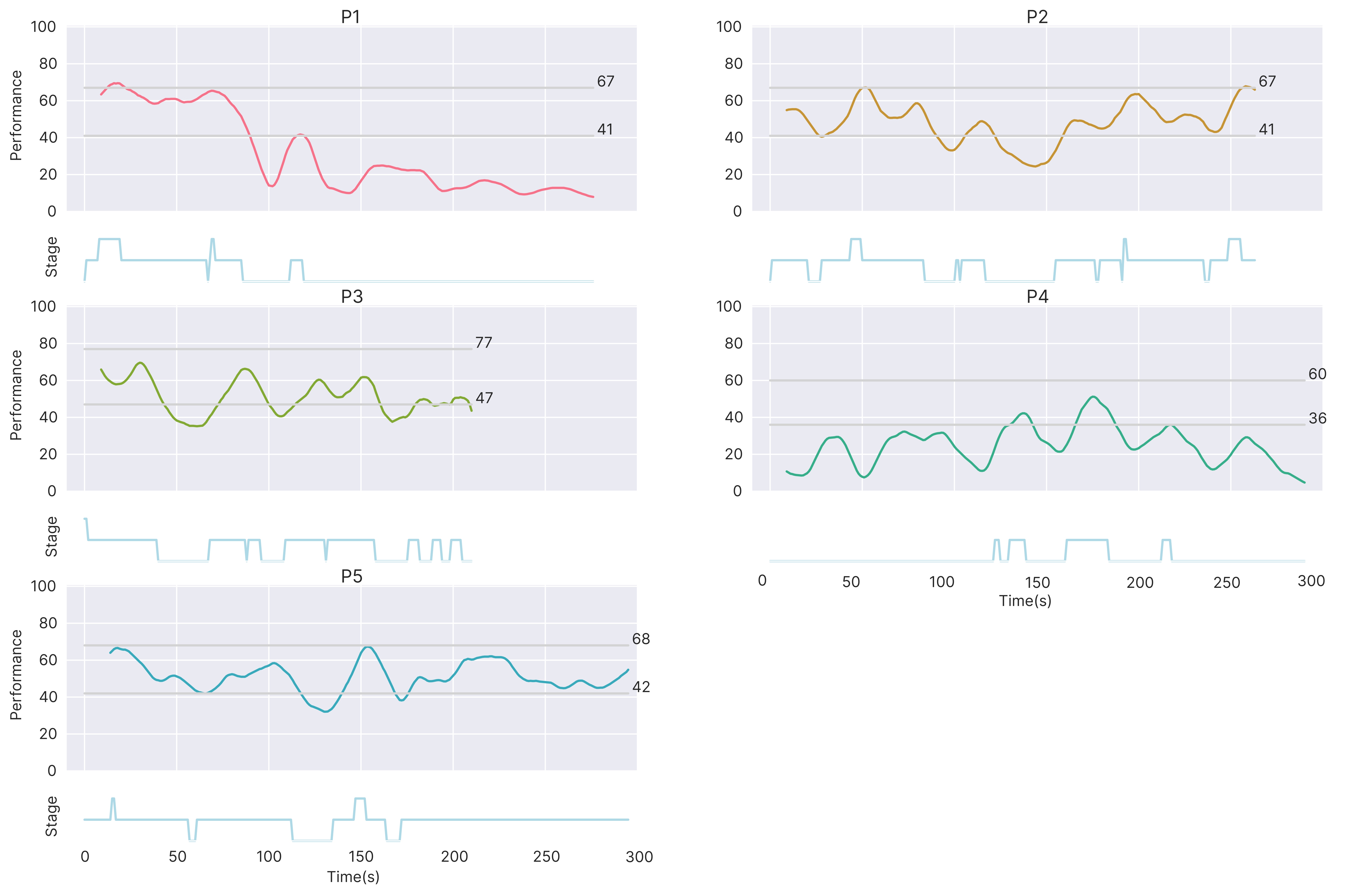}
    \vspace{-9mm}
    \caption{Social attention index logs of all autistic children in the field study, with their corresponding thresholds and performance stage switches in the session.}
    \label{fig:all_performance_roll10}
\end{figure}

In total, this study collected 1932 seconds of performance logs from five children and 8611 events (\eg, switching stages, laying golden eggs) using \name{}.
The children completed the game task (\ie, collecting two carts of eggs) in about 4-5 minutes, which was not too long to bore them or too short to achieve the effect of NFT.
\autoref{fig:all_performance_roll10} shows social attention index logs of the children with their respective personal thresholds for the performance stages as well as stage shifts along the time. 
Because of the noise in EEG raw signals, we smoothed the data with a moving average of a 10-second window.
The children were mostly in their medium performance stage (\ie, 28\%, 68\%, 62\%, 14\%, and 85\% of the whole gameplay time for P1-5, respectively).  
They jumped up and down between the three stages associated with the designed feedback, on average, 23 stage changes (11 up-switchings, 12 down-switchings).  
Such a pattern shows that the adaptive threshold has been effective, since to most of the children, this game was neither too hard nor too easy. 
They were able to remain on the medium performance stage while trying to achieve a higher performance.
Thus, children with different levels of social ability could be flexibly supported.
We observed that P1 performed worse in the later part of the session. 
By checking the video recordings, we found that P1 showed positive indicators during gameplay.
He was sometimes pointing to the tablet with a smile (\autoref{fig:field-study}(d)) and happily holding his father's hands at the end. 
However, a headband adjustment in the middle due to technical errors interrupted P1's focus and affected the performance later.
We also observed that P4 was mostly at the low-performance stage.
We thus checked the video recording and interaction logs and found that P4 showed a limited attention span. He performed well during the calibration stage when determining the thresholds for the game session, but lost focus quickly afterward.
To address that, the thresholds can be set manually to benefit autistic children at different levels.

\subsubsection{Interview Results.} \label{sec:qualitative-results}
A thematic analysis was conducted to analyze the transcribed data gathered from the semi-structured interviews with the participants.
The goal was to establish a set of structured, systematic understandings about how feedback, customization, and AR affect the user experience with \name{}.
In this section, we discuss our findings with the following themes, in light of the previous quantitative results.

\textbf{\name{} is mobile and convenient (D1).}
Overall, the participants appreciated the ease of setting up the game and were willing to play \name{} in a home setting. 
For instance,T6 noted that the game's set-up is quick and user-friendly, with the added advantage of using brain activities to adjust game elements, which can be helpful for children's social attention. T6 also believes that being able to use NFT games at home would greatly benefit families with autistic children, especially during COVID quarantine, by allowing for convenient and controlled training frequency and duration for the best outcome.

While home-setting NFT games are welcomed, \pqt{Parents may consider potential distractions from the environment that may affect children during the gameplay.}{T1}.
However, home-setting could make the game more enjoyable with more ways of playing. T4 envisioned that the parent and child could each play one game character to help each other, such as the child playing as the bird and the parent catching and handing over the egg.

\textbf{Feedback mechanisms enhance children's focus and engagement (D2).}
During the study, various positive behavioral signals were observed to indicate that the designed feedback of \name{} could engage and encourage the children. 
These signals were based on our observations and comments from the caregivers/guardians.
For instance, \autoref{fig:field-study}(e) shows P1 cheering up for the in-game character with thrilling emotion, gestures, and utterances.
P1 was extremely focused on the bird's movements, displaying unprecedented excitement and attention, as T1 commented.
Also, we found that P3 focused on the game all the time, sometimes pointed to specific game elements, and tried to mimic the body movements of the characters. 
T6 expressed that the background music tempo was engaging: \qt{I noticed him (P5) making sound by following the music to show his interest.}

However, T3 mentioned autistic children's attention might be different from typical children; \qt{They may not be able to notice all or fully comprehend the meaning of each feedback at the same time, especially for low-functioning autistic children.} 
Thus, obvious feedback like the bird's height and the colors of the eggs seemed to be most noticeable to the children. 
This implies the challenges in designing the feedback in NFT games.
T6 suggested that progressively, the caregiver/guardian could guide the child to explore other more subtle feedback (\ie, facial expression, sound effects) over a prolonged period of time.
Moreover, T4 mentioned that children would be more engaged if they are already familiar with the social interactions (\eg, turn-taking, high-five, shaking hands) in the game; thus, some social interactions in daily life could be used to improve the overall NFT outcomes.

\textbf{Design of AR and 3D benefits children in NFT (D3).}
From the participants' comments, we learned several advantages of using 3D and AR in \name{}.
First, 3D graphics could enhance the user experience as it is more attractive compared to 2D NFT games. 
T4 mentioned the scenes can attract more attention from children and improve the effectiveness of NFT when they are closer to reality.
Second, the above effect could be magnified by AR, which bridges the virtual game scenes to reality, helping children build the connection between the social content in the game and the real-life experience.
T2 echoed, \qt{This would be more effective [for NFT], because the background scenes are more realistic.}
T3 also commented, \qt{With 3D and AR, the kids know that such social collaboration and elements exist in the real world, so they can easily generalize the behaviors of the characters.}
T4 made a very interesting point, \qt{In the game, the characters are collecting and handing over eggs. The AR and 3D design could better encourage the child to mimic such social interactions, for example, gathering and passing apples to their Mom and Dad.}
Third, compared to traditional NFT games, AR-enhanced games could improve children's flexibility in thinking.
T6 notes that repeating activities in the same location can worsen over-routinized behaviors in autistic children. Using AR with different backgrounds can promote proactive thinking by helping children understand that social collaboration can occur in various situations and places.

Despite many benefits, T4 pointed out that the more realistic 3D elements and AR scenes might increase the likelihood of children touching and interacting with the game characters, instead of focusing on the game task.
However, this opens future opportunities for integrating AR with bodily games for ASD treatment.

\textbf{Game customization improves enjoyment (D4).}
\label{sec:qualitative-results-D4}
To users, one visible customization in \name{} is the process of coloring 2D characters on a paper sheet and later transforming them into 3D in-game models.
Participants thought this process was enjoyable and increased children's engagement.
T4 explained that integrating children's personal preferences into the game made it much more attractive.
T6 mentioned that many autistic children enjoy drawing and coloring, and seeing their drawings applied to in-game characters helps them build empathy. Additionally, T6 noted that this approach is a good opportunity to use the children's sense of space and imagination to connect coloring with 3D characters.
T3 mentioned \qt{He (P3) seems to like green things and wants everything to be green,} and freely coloring the characters allowed P3 to \qt{enjoy the game much more.} 
Free sketching of the characters and the customization of other game elements were suggested by the participants.
For instance, \pqt{Drawing an egg with their favorite color would allow them to pay more attention to the eggs and expect the eggs more. It will enhance their engagement.}{T2}.

On the contrary, T2 expressed that it might be hard for certain autistic children (\eg, low-functioning) to recognize characters that are actually from their colorings. In addition, some children may have less interest in drawing.
Thus, simplifying the coloring process for certain autistic children could address this issue, for example, by allowing children to drag and drop colors to different parts of the characters' costumes on the tablet.
Another customization in \name{}, the adaptive thresholds, could not be easily noticed by the participants. However, all the participants thought that the game difficulty was suitable for the children and that the NFT session was smooth.

\textbf{Comparing with previous experiences.}
In our interviews, we also asked the caregivers/guardians to compare the experiences of \name{} and Starkids \cite{Yang2021-cw} that has been used at the special education center for NFT, based on their knowledge and observation of the children.
We identified differences in two big categories: game effectiveness and engagement, which are detailed in \autoref{appendix:comparison}. 
Compared to Starkids, they stated that \name{} has a more effective design in terms of better game story introduction (\ie, via cartoon video), generalizable social actions (\eg, heads up, handing, placement), richer levels of feedback (\eg, changes of actions speed and facial expressions, reinforcer) yet avoid potential distraction (\eg, non-social actions).
Moreover, they thought that \name{} exhibits more engagement by leveraging customization, AR, and multilevel feedback.
First, AR in \name{} provides an immersive scene for better attentional focus and \pqt{helps children to experience and understand the object from different angles of observation in space.}{T1}.
Second, \name{} increases children's motivation via customization and gains engagement, whereas Starkids is barely customizable. 
Third, \name{} integrates game information into the social scene as multilevel feedback for both children and caregivers; Starkids directly uses non-social actions, progress bar, etc., leading to potential distractions.
However, the gaming theme of \name{} is limited, whereas Starkids offers a series of animal themes for children to choose from.

\section{Deployment Study} \label{sec:longterm-study}
Our field study targeted understanding the children's experiences with the novel features proposed in \name{}. However, NFT usually requires a long-term commitment to take effect.
We thus further conducted a three-week deployment study with P1 to understand children's experience in a slightly longer-term practice.

\subsection{Study Setup}
From our field study described above, P1 (a boy less than three years old) and his family agreed to participate in the deployment study. 
Besides the session in the field study, P1 took seven additional sessions in three weeks, and each study session was spaced two or three days apart. 
The additional study sessions had the same setup and procedure as above, which were integrated into P1's typical NFT therapy at the special education center (\autoref{fig:field-study}(f)).
For each session, we collected video recordings, system logs, and informal feedback from the caregiver and the guardian. 
At the end of the deployment study, we conducted semi-structured interviews with the caregiver and the guardian of the child to collect additional comments.

\subsection{Results}
In this section, we report the results of our deployment study, comprising the interaction logs of P1 and qualitative feedback from T1 and T5.

\begin{figure}[tb]
    \includegraphics[width=\linewidth]{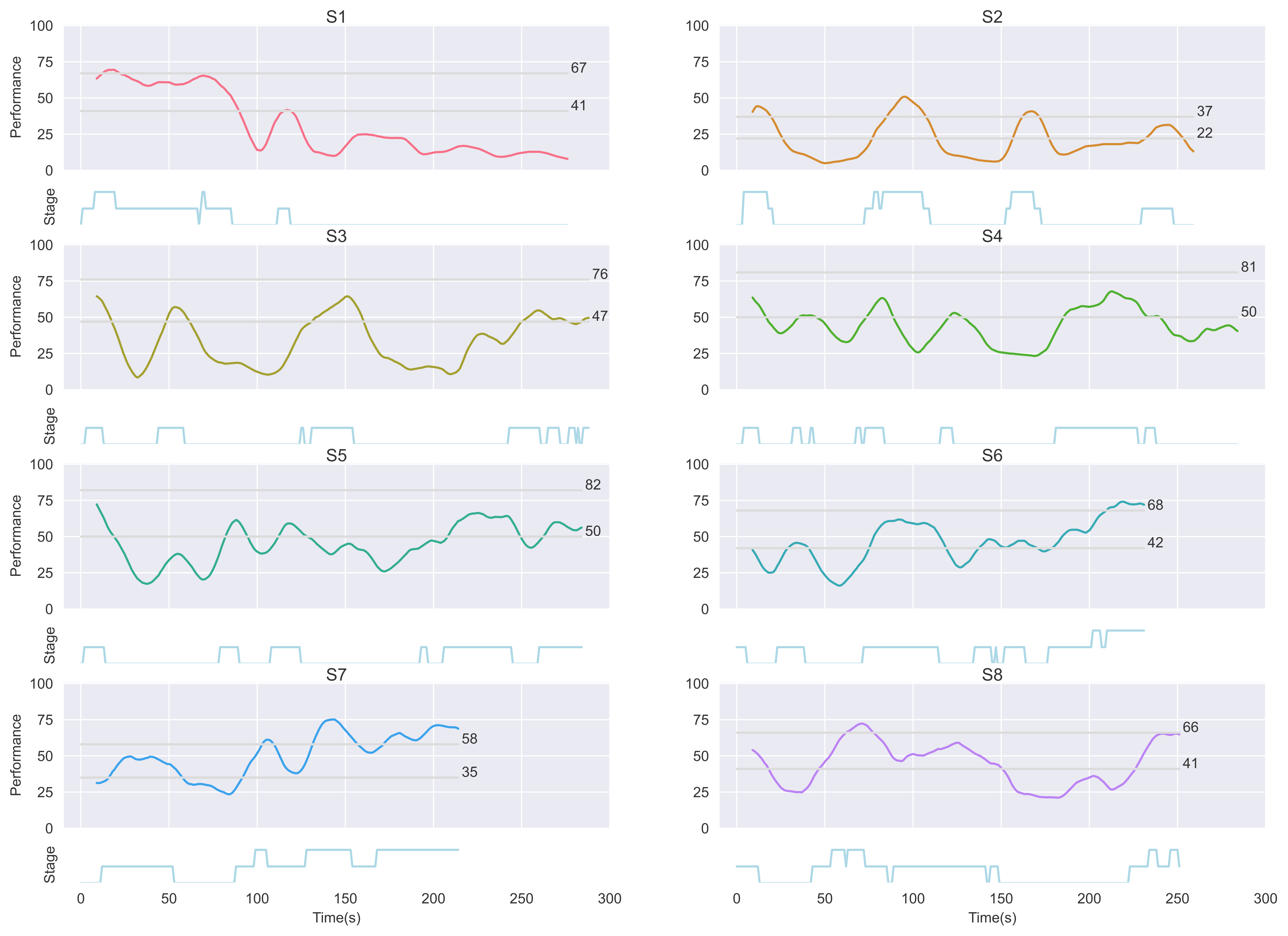}
    \vspace{-9mm}
    \caption{Social attention index logs of P1 with the corresponding thresholds of performance stages over eight sessions during the field study (S1) and the three-week deployment study (S2-7), ordered chronologically. }
    \label{fig:jinlin}
\end{figure}

\begin{table*}[tb]
    \centering
    \small
    \caption{Measures of P1's performance in eight playing sessions of the field study (S1) and the deployment study (S2-8).}
    \label{tab:jinlin}
    \vspace{-3mm}
    
\resizebox{\textwidth}{!}{%
    \begin{tabular}{ll|rr|rrr|rrr}
    \toprule
     \multicolumn{2}{c|}{\textbf{Measures}} & \multicolumn{8}{c}{\textbf{Sessions (ordered chronologically)}} \\
     \multicolumn{2}{c|}{\textbf{}} & \multicolumn{2}{c|}{\textbf{Week 1 (S1-2)}} &  \multicolumn{3}{c|}{\textbf{Week 2 (S3-5)}} & \multicolumn{3}{c}{\textbf{Week 3 (S6-8)}}\\ 
    \midrule
     \multirow{4}{*}{Performance Stage} & Thresholds $[t_1, t_2]$ & [41, 67] & [22, 37] & [47, 76] &[50, 81] & [50, 82] & [42, 68] & [35, 58] &[41, 66]\\
                & Low (\% time) & 67\% & 65\% & 71\% & 68\% & 62\% & 38\% & 22\% & 43\%\\
              & Medium (\% time) & 28\% & 15\% & 29\% & 32\% & 38\% & 50\% & 41\% & 46\%\\
              & High (\% time) &  5\% & 20\% & 0\% & 0\% & 0\% & 12\% & 37\% & 22\%\\
    \midrule
      \multirow{2}{*}{Social Attention Index} & Mean & 31.22 & 21.43 & 34.37 & 44.32 & 45.76 & 46.60 & 50.47 & 44.67\\
                                  & SD   & 21.91 & 13.49 & 17.72 & 12.80 & 14.49 & 15.51 & 15.38 & 15.59\\
    \bottomrule
    \end{tabular}
}
\end{table*}

\subsubsection{Interaction Logs}
During the study, we collected P1's data in seven additional NFT sessions, including 2724 seconds of social attention index logs and 12348 game events.
To analyze long-term effects, we combined the results of the seven sessions with that of the field study, denoted as S1-8, as shown in \autoref{fig:jinlin} and \autoref{tab:jinlin}.
\autoref{fig:jinlin} shows P1's social attention index values over time for each session. 
In Week 1 (S1-2), the overall social attend index was low, also with lower thresholds. 
In Week 2 (S3-5), while with much higher thresholds, P1 still stayed in the low-performance stage for long periods of time.
In Week 3 (S6-8), P1 was able to finish the game task faster, with stabler performance and more appropriate thresholds.
\autoref{tab:jinlin} provides quantitative measures of P1's performance.
For Weeks 1 and 2, P1 spent more than 60\% of the session time in the low-performance stage, despite that Week 1's thresholds were low. 
In Week 2, the high thresholds can be attributed to P1's increased familiarity with the game and the setup in the calibration phase; however, there was a difficulty to break through, shown as 0\% time in high-performance stage. 
Week 3's performance was much stabler, with more even distribution of the time in different performance stages, leaving space for improvement while not discouraging. 
Overall, the up-trend of the mean social attention index over three weeks indicates \name{}'s effectiveness in improving P1's abilities via NFT with our designed feedback and game mechanisms.

\subsubsection{User Feedback}
Our interviews in this deployment study focused on the effects of different aspects of \name{} in a longer-term practice of NFT. 
First, the richness of the feedback in the game can benefit autistic children with a gradual introduction to the game elements and progressive training of their abilities.
Children's initial interests in the game are essential for successful NFT, as T5 stated. 
Since autistic children usually have constrained attentional abilities, 
it is an opportunity for parents to guide children step by step to pay attention to the different levels and modalities of feedback, T5 stated. According to T1, children should initially focus on the eggs and the bird, which are the most apparent to comprehend. As they become more familiar with the game and perform better, attention can be directed to how characters catch and handle the eggs, as well as their facial expressions.

Second, the AR and customization of \name{} could facilitate children's engagement with NFT for an extended period. 
The 2D NFT games that P1 played before have single-mode feedback and fixed game scene; \pqt{Children soon get bored after multiple plays, because it becomes a repeated task without challenges and surprise.}{T1}.
The changeable environmental elements enabled by AR were considered to afford meaningful variations in the game experience that could benefit children with ASD. 
For example, %
playing Eggly on different surfaces, the table and with a flower behind it, can strengthen children's imagination and encourage open thinking as T1 commented.

Third, the mobility and convenience of the game may open more opportunities to combine NFT with behavior training in children's everyday life.
T1 advocated again that \name{} would be easily set up in a home setting, where many interactions between children and their parents could happen; thus, families could enforce children's attention to the social collaboration in the game by frequently mimicking the scenarios, facial expressions, and gestures.
T5 shared the same view on enhancing the (social) skills children learned via NFT games. 
For example, %
to reinforce P1's understanding of handing-over actions, T5 asked him to put away building blocks as the farm girl did, which allowed him to practice the action and develop a sense of order as well as reinforce NFT.

\section{Discussion}

In this session, we discuss extra design considerations learned from our study results, limitations of our approach, and future directions to enhance the work.

\textbf{Consider functioning levels of ASD children in design.} 
Due to the diverse abilities of autistic children, various considerations need to be integrated into the game design. 
For example, we should consider \pqt{providing different social interactions in the game to match their levels of functioning and social expectations,}{T5} as our results indicate that children's abilities to interpret the game scene affected their performance.
For instance, \pqt{handover and taking turns are hard for him [as a medium-functioning child] to understand. A simpler action like walking towards each other would be more appropriate to motivate him.}{T2}.
Moreover, we should consider an adaptive game duration for children's attention span and functioning level. 
We observed that high-functioning children could focus much longer than low-functioning ones, impacting their experience and NFT outcomes.

\textbf{Merge personal interests and preferences into game elements.}
First, offering a variety of game themes (\eg, cooking, marine, astronauts) and elements based on the children's interests could boost the NFT performance.
For example, some children are obsessed with cars, which could be leveraged to \pqt{display in the background as a reward to attract the child.}{T4}. 
This may require much more flexibility for the game design and development.
Second, more types of sensory stimulation, including visual, auditory, and haptic feedback, could be designed and tailored based on children's preferences.
Our interview results reveal that P3 and P4 prefer visual feedback while P2 and P5 prefer auditory feedback. 
P1 likes haptic feedback, which was not implemented in \name{}.
This diversity further poses challenges but also opportunities to create more versatile NFT games that allow for combining and customizing different sensory feedback elements.

\textbf{Connect game content to real-world social experience.}
The ultimate goal of NFT is help children generalize learned behaviors to their daily social scenarios.
Thus, it is critical to design game related to real-world practices for a better long-term outcome, such as \pqt{behaviors and interactions for exchange, cooperation, empathy, and accommodation, to support their social life.}{T2}.
Then, children could learn such social skills from the game and practice with their caregivers and family members. 
Moreover, T6 mentioned that social scenes with real human acting could be effective for children to acquire and use such skills. 
AR can be leveraged to put real humans in virtual backgrounds to demonstrate the actions applicable in different situations. 
This unique insight uses the opposite design idea in \name{}, which is interesting to explore further.

\textbf{Limitations and future work.}
While \name{} has shown effectiveness, and our studies have provided many deep insights into AR NFT, our approach has several limitations.
First, it is challenging to precisely investigate the children's attention and behaviors during gameplay without wearable devices such as eye trackers. 
We relied on our observations and the feedback from the caregivers/guardians, which may not be as accurate as quantitative measures.
Our study indicates that caregivers/guardians play an important role in facilitating children.
As a result, we argue that when designing NFT games, both the children and facilitators should be taken into consideration.
On the other hand, how caregivers/guardians affect the children's performance as a potential confounding factor, could be further studied to better understand how to achieve optimal training effects (\eg, how we may intentionally design to leverage the supportive effects from facilitators).
\rev{Moreover, a controlled experiment comparing AR and non-AR versions of the game can further examine more nuanced factors in the design.}
Thus, future studies may need to augment our current setup to obtain more insights.

Second, %
although the sample size of our study follows the convention of similar studies \cite{Matson-treatments-ASD}, larger-scale studies are needed to generate better understandings of AR NFT games.
Due to the sensitivity and practical challenge of the research context, conventionally, HCI studies involving preschool children with ASD have a relatively smaller sample size, making it difficult to yield robust, generalizable results. 
Therefore, a longer-term deployment study with more participants should be carried out, as NFT usually requires much time commitment to take effect.

Third, to support the best user experience, we employed WOZ for transforming colored 2D characters into 3D models in \name{}, so advanced computer vision techniques should be further explored to make this happen in real time. 
To offer much flexibility, advanced generative machine learning models can be leveraged to create 3D models, background music, 2D images, and other game elements based on children's interests and their real-time social attention index, which is a promising direction to research.

Fourth, as discussed above, much customization is demanded in the game design, including game characters, sensory stimulation, duration, etc. 
Currently, \name{} only supports limited ways of customizing the game.
Incorporating additional customization for game themes and elements could further improve the gaming experience for autistic children, such as allowing for selecting the object that the characters are collaborating over, determining the main awards, and preparing a list of animals or characters for children who are not interested in drawing.

\section{Conclusion}
We have introduced \name{}, a mobile AR NFT game for training the social attentional functioning of ASD children with customization and enjoyment while offering convenience for use in non-clinical settings.
Using a portable EEG headband and a tablet, \name{} was developed through a ten-month co-design process with four experts. 
We also derived a set of design principles for creating NFT games and proposed a multi-level feedback design framework for guiding future NFT game development.
\name{}, as an embodiment of the proposed framework, employs AR to enhance children's engagement and experience as well as uses various visual and auditory effects to strengthen the feedback loop.
We assessed \name{} through two studies, one single-session field study and one multi-session deployment study, at a special education center with five autistic children, accompanied by their caregivers and guardians. 
Quantitative and qualitative results indicate the effectiveness of \name{} and offer deep insights into the design and development of mobile AR NFT games as a playful ASD intervention.

\section*{Acknowledgments}
We thank all the participants in our studies for their valuable feedback to improve the system. This work is supported in part by the Discovery Grant of the Natural Sciences and Engineering Research Council of Canada (NSERC), the AI for Social Good Grant of the Waterloo AI Institute, and the NSSFC Art Grant (22CG184).

\bibliographystyle{ACM-Reference-Format}
\bibliography{references.bib}

\appendix

\section{Experts' Insights and Comments} \label{appendix:insights}
\begin{enumerate}[label=\textbf{I\arabic*}]
\item Provide an easy-to-access, cost-effective, and flexible training environment to offer a safe and controlled setting for autistic children.
    \begin{itemize}
        \item \pqt{Normally the parents accompany their children to our center for training periodically, like weekly or twice a week; sometimes the grandparents come regarding the parents are busy at work.}{E4}
        \item \pqt{Having the game accessible on a tablet allows greater flexibility regarding when and where the training occurs. It's essential for busy families who may not have time for frequent in-person appointments.}{E1}
        \item \pqt{Devices like Ipad and laptops are commonly used among families with ASD children so an easy setup mobile game would be a good idea.}{E2}
    \end{itemize}
    \item Offer positive feedback (reward) on the performance gain rather than the absolute performance to accommodate the limited and wide-ranged social ability of autistic children.
        \begin{itemize}
            \item \pqt{Consider only providing positive feedback, no negative feedback.}{E1}
            \item \pqt{The absolute value of the social attention index is not so important. The positive feedback that comes with the increase of index is important.}{E2}
        \end{itemize}
        
\item Employ multimodal gaming feedback with the visual channel as the primary for its effectiveness to grab autistic children’s attention.
        \begin{itemize}
            \item \pqt{Children mostly rely on sound and images for feedback; and would focus more on immediate feedback, which should be positive.}{E2}
            \item \pqt{Auditory feedback could be short as possible or serve as sound effect to boost the social sense, to fit the potentially significant changes or fluctuations in performance.}{E2}
            \item \pqt{Most autistic children are visual learners; thus, visual cues are better supporting than auditory cues within the training.}{E1}
        \end{itemize}

\item Integrate the social attention index dynamically and smoothly into the game scene to make it easily understood and avoid distraction for autistic children.
        \begin{itemize}
        \item \pqt{The explicitly knowing of the mapping of the game helps users on their performance.}{E1}
         \item \pqt{The child needs to know what is good or bad feedback in order to learn how to control brain activity.}{E3}
         \item \pqt{The social attention index bar does not make much sense to the kids. They rely on sound and images for feedback. The bar is mainly for the caregiver’s reference. However, the movement of the bar may still distract the child during training.}{E2}
        \end{itemize}

\item Exhibit social collaboration actions in the game that would happen in real life for autistic children to imitate and apply in their daily lives.
    \begin{itemize}
        \item\pqt{The mirror neuron is effectively triggered when the child observes a social movement and can empathize with it.}{E2}
        \item \pqt{The social interaction in the game should be real, which means the interaction could actually happen in daily life so that kids can learn or copy and apply the skill to their real life.}{E1}
        \item \pqt{The social interaction is better to be a cooperation task among the characters through some direct body touch or indirectly via objects, like transferring a ball.}{E1}
         \item \pqt{Social interaction should be present throughout the game - even below the threshold.}{E2}
    \end{itemize}

\item Design human-like characters with apparent social gestures and body movements to allow autistic children to better connect to the real-world.
    \begin{itemize}
        \item \pqt{Apparent social gestures, such as looking into each other and shaking hands, are essential for the success of NFT games regarding social skills training.}{E3}
        \item \pqt{Autistic children will have more attention to objects rather than lives like humans or animals. We hope the game design can lead kids to be interested in humans, which is essential in training social skills. Thus, humanoid characters are desired.}{E1}
    \end{itemize}

\item Provide interactive and less-realistic facial expressions to combat the difficulty of autistic children in recognizing and imitating them.
\begin{itemize}
    \item \pqt{It would be nice if the character could change their facial expression. Also, the face should not be super realistic.}{E3}
    \item \pqt{Provide an opportunity for children to match correct facial expressions with interactions and to learn and imitate the emotional response.}{E3}
\end{itemize}

\item Exclude textual dialogue in the scene as it interferes with the trigger of mirror neurons of autistic children.
    \begin{itemize}
        \item \pqt{The social interaction should not be verbal communication.}{E2}
    \end{itemize}

\item Indicate the performance levels via social interactions in the game for keeping engaging autistic children.
        \begin{itemize}
         \item \pqt{Autistic children have more interested in interacting with humanoid robots due to the highly predictable nature of the programmed actions. Correspondingly, the programmed actions of the characters, like body movement and facial expressions, could change during the game over some rules, like different stages of performance.}{E1}
        \item \pqt{Staged feedback using discrete variables, like different dance positions, or non-staged feedback using continuous variables, like different rates of movement, the design shall depend on the game scene.}{E2}
        \end{itemize}
\item Indicate the gameplay progress in the designed social scene to ease the expectation from autistic children.
        \begin{itemize}
        \item \pqt{The child may feel anxiety if they do not know when the game will stop. }{E1}
        \item \pqt{The movement of the progress bar in existing games may distract the child during training. To minimize the distraction, embedding the progress bar into the main scene close to the characters could be a good way.}{E2}
        \end{itemize}
        
\item (Re)enforce the positive feedback to keep engaging and interesting autistic children based on their unique sensory and attentional profiles.
        \begin{itemize}
        \item \pqt{Children have limited interests. To interest the child more in the game by bringing the child's personal interests and familiar objects as the reinforcer - an effective reward. Some objects like vehicles and blowing bubbles commonly interest children.}{E1}
        \item \pqt{The reinforcer can be used as an external reward, which could not be as part of the social actions.}{E1} 
        \item \pqt{If the character could say something like “thank you” would be good social training for the kids, so does look into others’ eyes.}{E3}
        \end{itemize}

\item Leverage AR and 3D graphics to enhance motivation and engagement as well as build real-world connections.
   \begin{itemize}
        \item \pqt{Children with ASD normally have difficulty with training. AR and 3D graphics could increase the participation and motivation to the NFT training, making the training more engaging and appealing to them.}{E2}
        \item \pqt{AR game potentially provides a contextual learning experience, where children can learn the actions in a real-world background. This help increase their ability to transfer their skills to real-life situations.}{E1}
        \item \pqt{ AR and 3D graphics motivate children with ASD to focus and observe the game scene in a controlled and familiar background.}{E2}
    \end{itemize}

\item Consider customization for children with different severity in ASD.
    \begin{itemize}
        \item \pqt{Current solutions do not consider the differences in severity among autistic children.}{E1}
    \end{itemize}
    
\item Bring children’s familiar/personal objects into the game to close the gap between the virtual and real worlds while providing variety across sessions.
    \begin{itemize}
        \item \pqt{To interest the child more into the game by bringing the personal interests and familiar objects of the child and having the variety among different sessions.}{E3}
        \item \pqt{The familiarity with the object (reinforcer) helps children engage in the game.}{E2}
        \end{itemize}
\end{enumerate}

\section{Comparison between \name{} and Starkids}\label{appendix:comparison}

\textbf{Effectiveness}
\begin{itemize}
\item Story Introduction
    \begin{itemize}
        \item \name{}: Cartoon video
        \item Starkids: Description
    \end{itemize}
    
    \pqt{The introduction is for children to understand the social story and tasks they are about to watch. Using animated videos to convey the story could be easier for children to understand compared to verbal description.}{T4}

\item Social Movements Design
    \begin{itemize}
        \item \name{}: Heads up, hands up \& down, catching, turning, handing, placement
        \item Starkids: Dancing, cheering with hands up
     \end{itemize}

     \pqt{To trigger mirror neuron activity, it's important to choose actions that children can understand. In addition to key social interactions and feedback, non-social in-game actions can be a distraction. \name{}'s social actions are easy for kids to understand. The cheering as an action aside from the social scene in Starkids could be a distraction, which is supposed to be designed as a reward.}{T1}

\item Feedback Design
    \begin{itemize}
        \item \name{}: Change of movement speed and facial expression (Storytelling feedback)
        \item Starkids: Change the position of snake \& Audio speed
    \end{itemize}

    \pqt{\name{} provides richer feedback, like the colorful eggs and the bubbles and there seem to be various factors changing and other game elements appearing. Starkids's feedback is simpler, the change of speed and dancing position. Feedback diversity can determine the level of fun in the game, making children feel more varied and not so monotonous.}{T1}

\item Graphic shape
    \begin{itemize}
        \item \name{}: 3D
        \item Starkids: 2D
     \end{itemize}
     \pqt{The social interactions in games also serve as a demonstration and hopefully initiate the imitation of children in the real world. We hope children imitate the actions demonstrated in Starkids, however, 2D abstracts the actions making it difficult for children to imitate. \name{} promotes 3D, making the animation and movements more accurate and clear for children to observe and imitate.}{T4}
\end{itemize}

\textbf{Engagement}
\begin{itemize}
   \item Customization
    \begin{itemize}
        \item \name{}: Customized characters
        \item Starkids: Not applicable
     \end{itemize}
     \pqt{Coloring as a way of customization definitely increases the motivation, because if there's an in-game character or role the child likes, he may be more proactive in participating.}{T4}

    \item AR
    \begin{itemize}
        \item \name{}: AR background
        \item Starkids: Not applicable
     \end{itemize}
     \pqt{\name{} has a simpler AR interface, which is easier for children to focus on with less distraction.}{T1}\pqt{It may also leverage the real-world background combined with the game, where children could learn various locations with more engagement.}{T1}
     
     \pqt{AR helps children to experience and understand the object from different angles of observation in space.}{T1}
     
     \pqt{I don't see any negative impact on AR; however, it may need imagination from kids to pay attention to the AR background.}{T4}

 \item Feedback Design
    \begin{itemize}
        \item \name{}: Embedded immediate \& progress feedback design for both children and caregivers 
        \item Starkids: Progress bar \& index bar
     \end{itemize}
     \pqt{Some children may be very focused on a point in a picture, for example, if a bar graph is constantly changing, it may become a focus of the child's attention. Children may get anxious and stop cooperating after reaching a certain point, as he doesn't know how long the game will last. Using part of the social scene rather than a standalone graph to display the information has the advantage of reducing the potential distraction from the graph.}{T1}
     
\item Themes
    \begin{itemize}
        \item \name{}: Single theme of farm
        \item Starkids: Series of animals themes
    \end{itemize}
    \pqt{I think \name{} could be enriched by having more themes.}{T1}
\end{itemize}

\end{document}